\newif\ifarxiv
\arxivtrue

\ifarxiv
\documentclass[a4paper,11pt,abstract=true]{scrartcl}
\usepackage{libertine}
\usepackage{natbib}
\usepackage{hyperref}
\newcommand\Description[1]{}
\usepackage[includeheadfoot, head=13pt, foot=2pc, top=58pt, bottom=44pt, inner=46pt, outer=46pt, marginparwidth=2pc, heightrounded]{geometry}
\else
\documentclass[format=acmsmall,review,natbib]{acmart}
\usepackage{libertine}
\fi

\usepackage{mathpazo}
\usepackage[T1]{fontenc}
\usepackage[utf8]{inputenc}
\usepackage[scaled=0.81]{beramono}  %

\usepackage[multiple]{footmisc} %
\usepackage{tabularx}
\usepackage[yyyymmdd]{datetime}

\ifarxiv\else
\makeatletter
\def\@copyrightpermission{}
\def\@copyrightowner{}
\acmPrice{}
\makeatother
\fi

\setcitestyle{square,numbers,sort}

\usepackage{graphicx}
\graphicspath{{.}{figs/}}
\DeclareGraphicsExtensions{.png,.pdf}

\usepackage{xparse}
\usepackage{dcolumn}
\usepackage{subcaption}
\renewcommand\paragraph[1]{\textbf{#1}}

\usepackage{xstring}

\usepackage{numprint}

\title{Applying Bayesian Analysis Guidelines to Empirical Software Engineering Data}
\subtitle{The Case of Programming Languages and Code Quality}

\ifarxiv
\author{Carlo A.\ Furia$^1$
        $\quad\cdot\quad$
        Richard Torkar$^{2,3}$
        $\quad\cdot\quad$
        Robert Feldt$^2$ \\[2mm]
        \normalsize
        $^1$ Software Institute, USI Universit\`a della Svizzera Italiana, Switzerland \\
        \normalsize
        $^2$ Chalmers and the University of Gothenburg, Sweden \\
        \normalsize
        $^3$ Stellenbosch Institute for Advanced Study (STIAS), South Africa
}
\else
\author{Carlo A.\ Furia}
\orcid{http://orcid.org/0000-0003-1040-3201}
\affiliation{%
  \institution{Software Institute, USI Universit\`a della Svizzera italiana}
  \city{Lugano}
  \country{Switzerland}}
\email{https://bugcounting.net}

\author{Richard Torkar}
\orcid{ORCID}
\affiliation{%
  \institution{University of Gothenburg}
  \city{Gothenburg}
  \country{Sweden}}
\affiliation{%
  \institution{Stellenbosch Institute for Advanced Study (STIAS)}
  \city{Stellenbosch}
  \country{South Africa}}
\email{https://www.torkar.se}

\author{Robert Feldt}
\orcid{ORCID}
\affiliation{%
  \institution{Chalmers and University of Gothenburg}
  \city{Gothenburg}
  \country{Sweden}}
\email{https://www.roberfeldt.net}
\fi

\ifarxiv\else
\authorsaddresses{}

\acmBooktitle{}
\fi

\date{\today} %

\makeatletter
\let\Title\@title
\makeatother

\usepackage{amsmath,amsthm}

\theoremstyle{definition}

\usepackage{amsfonts}

\newcommand{\model}{\ensuremath{\mathcal{M}}}

\newcommand{\lcal}{\ensuremath{\mathcal{L}}}

\usepackage{proof}

\usepackage{xstring}
\DeclareRobustCommand{\var}[1]{\ensuremath{\textnormal{\textsl{#1}}}}
\DeclareRobustCommand{\shortvar}[1]{\ensuremath{\textnormal{\textsc{\StrLeft{#1}{1}}}}}
\DeclareRobustCommand{\coef}[1]{\ensuremath{\beta^{\shortvar{#1}}}}

\DeclareRobustCommand{\dist}[1]{\ensuremath{\textsf{#1}}}

\usepackage{booktabs}
\usepackage{xcolor}
\definecolor{acol}{RGB}{27,158,119}
\definecolor{bcol}{RGB}{217,95,2}
\definecolor{ccol}{RGB}{117,112,179}
\definecolor{dcol}{RGB}{166,216,84}
\definecolor{ecol}{RGB}{231,138,195}

\usepackage{rotating}
\usepackage{multirow}

\usepackage{enumerate}
\usepackage[inline]{enumitem}
\setlist[description]{leftmargin=\parindent,labelindent=\parindent,align=left}

\usepackage{changepage}

\usepackage{tikz}
\tikzstyle{lang}=[draw=none,font=\footnotesize,inner sep=1pt,outer sep=1pt]
\usetikzlibrary{positioning,shapes.geometric,arrows,calc}
\usepackage{pgfplots}

\usepackage{array}

\usepackage{xspace}

\usepackage{xstring}

\usepackage{ifthen}

\usepackage{fontawesome}
\definecolor{ncgreen}{RGB}{102,194,165}
\definecolor{ncorange}{RGB}{252,141,98}
\definecolor{ncviolet}{RGB}{141,160,203}
\newcommand{\checkOK}[1][ncgreen]{\text{\color{#1}\faCheck}}
\newcommand{\checkFail}[1][ncorange]{\text{\color{#1}\faClose}}

\pgfplotsset{compat=1.16}

\begin{document}

\ifarxiv
\maketitle
\fi

\begin{abstract}
  Statistical analysis is the tool of choice to turn data into information, and then  information into empirical knowledge. The process that goes from data to knowledge is, however, long, uncertain, and riddled with pitfalls. To be valid, it should be supported by detailed, rigorous guidelines, which help ferret out issues with the data or model, and lead to qualified results that strike a reasonable balance between generality and practical relevance. Such guidelines are being developed by statisticians to support the latest techniques for \emph{Bayesian} data analysis. In this article, we frame these guidelines in a way that is apt to empirical research in software engineering.

    To demonstrate the guidelines in practice, we apply them to reanalyze a GitHub dataset about code quality in different programming languages. The dataset's original analysis~(\citeauthor{FSE}, \citeyear{FSE}) and a critical reanalysis~(\citeauthor{TOPLAS}, \citeyear{TOPLAS}) have attracted considerable attention---in no small part because they target a topic (the impact of different programming languages) on which strong opinions abound.
    The goals of our reanalysis are largely orthogonal to this previous work, as we are concerned with demonstrating, on data in an interesting domain, how to build a principled Bayesian data analysis and to showcase its benefits. %
    In the process, we will also shed light on some critical aspects of the analyzed data and of the relationship between programming languages and code quality---such as the
    impact of project-specific characteristics other than the used programming language.

    The high-level conclusions of our exercise will be that Bayesian statistical techniques can be applied to analyze software engineering data in a way that is principled, flexible, and leads to convincing results that inform the state of the art while highlighting the boundaries of its validity.
    The guidelines can support building solid statistical analyses and connecting their results, and hence help buttress continued progress in empirical software engineering research.
  \end{abstract}

\ifarxiv\else
\maketitle
\fi

\section{Introduction}
\label{sec:introduction}

Empirical disciplines, including a substantial part of software engineering research,
mine data for information, and then use the information as evidence
to build, extend, and refine empirical knowledge.
Statistical analysis is key to implementing this process;
but statistical techniques are just tools, which need 
detailed \emph{guidelines} to be applied properly and consistently.
It is only through the combination of powerful statistical techniques\
and rigorous guidelines to apply them
that we can distill empirical knowledge following a process
that is consistent, rests on solid principles, and
ultimately is more likely to lead to valid results
with a higher degree of confidence.

Whereas frequentist statistical techniques have been commonplace in science 
for over a century---since the influential work of the likes of Pearson~\cite{Pearson-work} and Fisher~\cite{Fisher-work}---the
state of the art in applied statistics is moving towards 
using \emph{Bayesian} analysis techniques.
As we discussed in previous work~\cite{FFT-TSE19-Bayes2},
recent developments in Bayesian analysis techniques (such as using Hamiltonian Monte Carlo fitting algorithms~\cite{brooks2011handbook})
coupled with an increasing availability of the computing power needed to run them
on large datasets
have convincingly demonstrated the advantages of using Bayesian statistics
and the flexibility and rigor of the analysis they support.
More recently, applied statisticians have also been working out
practical \emph{guidelines} that can boost usability
and impact of Bayesian statistical data analysis~\cite{aczel20guidelines,schadBV20workflow,gabry2019visualization,gelman20workflow}.
In this paper, we present some of these guidelines and frame them
in a way that is suitable for empirical research in the software engineering domain---with
the goal of demonstrating how they can support a principled way of building
statistical analyses of software engineering data.

To demonstrate the guidelines in practice, we follow them to analyze
a large dataset about the code quality 
of projects written in disparate programming languages 
and hosted on GitHub~\cite{FSE}.
The empirical study that curated this dataset and performed the original analysis~\cite{FSE}
was followed by a critical reanalysis by a different group of researchers~\cite{TOPLAS};
as we recall in \autoref{sec:FSE-TOSEM},
the topic has received much attention and stirred some controversy.
This visibility makes the dataset an attractive target for our own purposes.
 
In the paper, %
we go through various aspects of the data analysis performed in 
the previous studies~\cite{FSE,TOPLAS},
illustrating the versatile features of Bayesian statistical models in practice.
We demonstrate how the guidelines
support an incremental and iterative analysis process, where several
key features of a statistical model can be validated; this, in turn, 
encourages trying out different models and comparing them
in a rigorous way---as opposed to blindly relying on one-size-fits-all rules of thumb.
Following this process, we demonstrate that some issues of the original analysis~\cite{FSE} or criticized by the follow-up reanalysis~\cite{TOPLAS}
could have been identified more easily.
Furthermore, the limitations and actual impact of previous studies %
could have been framed more straightforwardly and more transparently.
The conclusion of our exercise will be that
flexible statistical techniques coupled with principled and structured guidelines 
can help address empirical research questions directly and
transparently. This can lead to explanations
that are nuanced and detailed, and hence, ultimately, 
that can become convincing foundations for building shared knowledge.

\subsection{Dataset and previous studies}
\label{sec:FSE-TOSEM}

In this paper, we reuse the dataset collected and analyzed by \citeauthor{FSE} 
in a paper published at the FSE\footnote{%
  The ACM SIGSOFT International Symposium on Foundations of Software Engineering.%
} conference~\cite{FSE}.
A critical reproduction~\cite{TOPLAS} of the original study, written by \citeauthor{TOPLAS} and published in the TOPLAS\footnote{%
  The ACM Transactions on Programming Languages and Systems. %
}
journal, triggered a prolonged controversy that reverberated on social media.

Here is the story so far, in the shortest terms possible:\footnote{Hillel Wayne provides a much more detailed account \url{https://www.hillelwayne.com/post/this-is-how-science-happens/}.}
based on their analysis of projects hosted by GitHub,
the original study~\cite{FSE} (henceforth, ``FSE'') claimed to have found an association between certain programming languages and the bug proneness of code written in them.
The reproduction study~\cite{TOPLAS} (henceforth, ``TOPLAS'')
criticized several aspects of FSE---most prominently, its data collection and classification practices---and
questioned the soundness of some of its results.
In a rebuttal, 
the authors of FSE defended their results;\footnote{\url{https://arxiv.org/abs/1911.07393}}
and in a rebuttal of the rebuttal the authors of TOPLAS maintained their criticism.\footnote{\url{http://janvitek.org/var/rebuttal-rebuttal.pdf}}

Our paper is \emph{emphatically \textbf{not}} our attempt to jump into the fray:
we do not have much to add to the subject matter of the controversy.
However, we appreciate the interest that the controversial topic received,
and see it as an opportunity to present our views on a different, but related, aspect:
practices in statistical analysis.
Both FSE and TOPLAS primarily use \emph{frequentist} statistical techniques.
Even in the best conditions, these techniques' flexibility is limited in comparison to the \emph{Bayesian}
statistical techniques we have been advocating~\cite{FFT-TSE19-Bayes2,TFFGGLE-PracticalSignificance}.

Overall, our contributions fall largely outside the focus of FSE and TOPLAS---except to the extent that they target the
same domain and the same data. The core of both papers revolves around GitHub data,
how it was collected and processed,
and how the variables
of interest have been operationalized. TOPLAS's main goal was to attempt to reproduce FSE's results,
and hence it deliberately %
makes mostly limited changes to the statistical models.
Our analysis 
takes the data as it was collected and made available
by FSE's original study,
and tries to make the most out of it
following rigorous guidelines to apply flexible statistical practices---Bayesian statistics, that is.

\subsection{Overview}

The overall goal of this paper is
demonstrating how Bayesian statistical techniques
can be applied in a principled way
to build suitable statistical models.
These models can then be used to 
answer research questions in a flexible\footnote{%
  Here, ``flexible'' means that it
  can be adapted to different kinds of scientific questions and analysis domains 
  while remaining effective. %
  } way,
and to quantify limitations and uncertainties
about what one can reliably infer from the models.

This complements our earlier work on Bayesian data analysis for empirical software engineering~\cite{FFT-TSE19-Bayes2,TFFGGLE-PracticalSignificance},
which:
\begin{itemize}
    \item Argued for using Bayesian over frequentist statistics
    and showcased the former's flexibility on software engineering data~\cite{FFT-TSE19-Bayes2}.
    
    \item Suggested to analyze \emph{practical} significance using a combination
    of Bayesian statistics and cumulative prospect theory,
    which helps stakeholders evaluate the impact of a technique or a practice
    in a way that takes into account their constraints, available resources, and intuitive reasoning~\cite{TFFGGLE-PracticalSignificance}.
\end{itemize}

\subsubsection{Key benefits of Bayesian statistics}

Before we go into the novel contributions of the present paper,
let us briefly summarize the benefits of Bayesian statistical techniques---which
we presented in detail in our previous work~\cite{FFT-TSE19-Bayes2,TFFGGLE-PracticalSignificance}.
\autoref{sec:bda-results} will further demonstrate several of these benefits on the programming language case study.

There is a growing awareness in several empirical scientific disciplines
that ``classical [frequentist]
statistical tools are not diverse enough to handle many common research questions''~\cite{mcelreath2020statistical}.
Bayesian statistics, in contrast, are much more flexible,
as they provide general methods to connect data, models,
and research questions~\cite{Bayes-whatsabout}.
Bayesian statistics are more flexible
because they are centered around \emph{modeling}:
how we fit a Bayesian model is largely independent of the details
of how the model was built.
In contrast, different frequentist models often require widely
different analysis procedures: if we need to tweak the model or
its underlying assumptions even slightly, the frequentist analysis results
may become unreliable.

The main output of a Bayesian data analysis is a \emph{distribution}
of model parameters fitted on the data.
This includes rich quantitative information,
in contrast to the point estimates that are the usual outcome
of applying frequentist statistics.
Providing distributional information is a key strength of Bayesian
statistics.
First, it supports quantitative and nuanced analyses instead
of a purely dichotomous (yes/no) view---which is prevalent
with statistical hypothesis testing (a core technique of frequentist
statistics that has been under intense scrutiny~\cite{ASA-statement,rise-up}).
Second, distributional information is easier to understand,
since it measures quantities of interest in the specific domain.
Contrast this to purely statistical metrics such as $p$-values or confidence intervals, which are notoriously hard to interpret correctly~\cite{misinterpret-pvals,misinterpret-cis}.
Third, the quantitative distributional information
that is provided by a fitted Bayesian model
supports simulating the \emph{derived} distributions
of a variety of quantities of interest, such as
outcomes in a specific scenario---which is especially useful to
analyze practical significance~\cite{TFFGGLE-PracticalSignificance}.

The current paper focuses on how to apply Bayesian data analysis
in a principled way: following detailed guidelines and a structured workflow.
Demonstrating the guidelines on FSE's dataset, 
our contributions address four 
aspects that are relevant
to every study that involves statistical data analysis.

\subsubsection{How to design a statistical model?}

Modeling requires to exercise \emph{judgement}---something
that can be based on practices, customs, and heuristics, but is not completely reducible to a fixed set of rigid rules.
Bayesian statistics emphasizes the \emph{modeling} aspect of data analysis,
and provides quantitative techniques to help ground heuristics and practices onto a robust and sound statistical framework.

\autoref{sec:guidelines} presents guidelines to build a Bayesian statistical
model incrementally (adding features as needed), iteratively (improving
a model based on the shortcomings of the previous ones), and
rigorously (with quantitative criteria to assess a model's suitability).
Our guidelines customize general guidelines developed by the
Bayesian data analysis community to the scenarios that are
common in empirical software engineering.
\autoref{sec:bda-pl} demonstrates, on the FSE dataset, that our guidelines 
provide principled ways of assessing the strengths and weaknesses of 
any statistical model for the analysis at hand. %

\subsubsection{How to spot data problems?}

TOPLAS's criticism of FSE's analysis questions %
the accuracy of some of the \emph{data} that was collected and how it was
processed. For example, it says that 
``project size, computed in the FSE paper as the sum of inserted lines, is not accurate---as it does not take deletions into account''~\cite[\S{}3.2]{TOPLAS}.
Can Bayesian statistical techniques help discover problems with the data---such
as inconsistencies, sparseness, and lack of homogeneity---that limit
the validity and generalizability of the statistical analysis's results?

Naturally, no statistical technique (no matter how powerful) can
supersede a careful analysis of construct validity~\cite{feldt2010validity,ralphT18construct}, which
should precede the statistical analysis and lay the foundations for it.
Still, applying the Bayesian guidelines that we present
can ferret out issues with the data and highlight
where uncertainty %
is more or less pronounced,
so that we can heed any limitations when drawing conclusions.
For example, \autoref{sec:data-problems} finds that
the number of inserted lines performs poorly as a predictor,
echoing TOPLAS's observation that it may not be a suitable measure of size.

\subsubsection{How to assess significant results?}

In previous work~\cite{FFT-TSE19-Bayes2}, we demonstrated
that Bayesian statistical techniques 
can help move away from a dichotomous (significant\slash not significant) framing of
research questions---which comes typically with
frequentist null hypothesis testing and is often artificially restrictive---and
instead focus on \emph{practical} significance~\cite{TFFGGLE-PracticalSignificance}.

The gap between statistical significance and practical significance is
more likely to be wide when studying complex domains with plenty of confounding factors.
The analysis of programming language data is
a clear example of such complex domains.
In this paper, %
we show how the %
Bayesian statistics guidelines
support a nuanced analysis of complex models,
and help keep the focus on concrete scenarios and practically relevant measures.
Concretely, we show that Bayesian data
analysis provides a flexible model of data distributions, which can be
used to \emph{predict} outcomes in different scenarios directly in
terms of statistics that are based on variables in the problem domain.

Our analysis's conclusion 
will be that the key question ``which programming languages
are more fault prone'' does not admit a simple straightforward answer---not with the analyzed data at least. 
Nevertheless, as we argued in~\cite{FFT-TSE19-Bayes2} and now demonstrate in \autoref{sec:practical-significance}, practitioners can ask specific questions and answer them by running simulations
on the Bayesian model, rather than having to rely on general results that may not be meaningful in their context.

\subsubsection{How to build knowledge incrementally?}
Every empirical study has limitations;
lifting them requires to perform new experiments.
Another advantage of Bayesian statistical models 
built using an incremental process is that they can be refined
as we collect more data. %
This way, our models become better over time since they accurately reflect %
the evolving scientific knowledge in a certain area.

\autoref{sec:follow-up} discusses 
how applying Bayesian analysis guidelines 
helps plan for additional data collection
based on the limitations of the analyzed data.
Different experiments are no longer merely a loose collection
around the same themes, but can be planned back-to-back 
in a way that progressively reduces the uncertainty in knowledge.

\subsection{Contributions}
\label{sec:contributions}

This paper makes the following contributions:

\begin{itemize}
    \item It presents guidelines to apply Bayesian statistics following
    a systematic process that goes from building and validating
    the model to fitting and analyzing it.
    
    \item It demonstrates the guidelines by showing how to
    incrementally build a suitable statistical model to capture FSE's
    language quality data.
    
    \item It analyzes the fitted model to investigate the original
    questions of the effect of programming languages on fault proneness
    with a focus on practical scenarios.
    
    \item For reproducibility, all analysis scripts are available online together with additional results and detailed data visualization:
    \begin{center}
        {\small \textsc{replication package:}}$\quad$ \url{https://doi.org/10.5281/zenodo.4472963}$\quad$ \cite{replication-package}.
    \end{center}
\end{itemize}

\subsubsection{Scope}
\label{sec:scope}

To a large degree, the guidelines we present are \emph{not specific}
to certain classes of statistical models or analysis domains.
The case study we detail in this paper is about programming language quality,
which we model using several generalized linear models of different complexity---a
broad class of statistical models widely used for their flexibility.
This does not mean that the guidelines are only applicable to programming language data,
nor that they only work for generalized linear models.

Since the guidelines are largely independent of the specific features of
the chosen statistical model, the domain-specific details of how to
operationalize a certain data analysis problem and how to
build a valid ``construct'' (a statistical model)
are largely outside the scope of the present paper.
As we remarked above, we selected the programming language data
as case study because it has been already thoroughly analyzed and scrutinized
(albeit in a frequentist setting);
thus, we can build on FSE's and TOPLAS's work to demonstrate
the additional steps to be taken to bolster the validity of a statistical data analysis.
This leaves room for \emph{different} analyses of the same research questions
but using different data collection processes or different statistical models.
Our guidelines remain valid as a safeguard against modeling mistakes or shortcomings;
since they promote an \emph{iterative} approach, they can also suggest what to change when
they fail to validate a candidate model.

\subsubsection{Organization}

The rest of the paper is organized as follows.
\autoref{sec:guidelines} illustrates Bayesian data analysis guidelines
with an angle that is relevant for empirical software engineering.
\autoref{sec:bda-pl} follows the guidelines
to incrementally build a model
that is suitable to capture FSE's programming language data.
Various models are rigorously evaluated and compared,
so that the final model is arguably the ``best'' among them according
to certain quantitative criteria.
\autoref{sec:bda-results} analyzes the fitted model
to study the original questions of which programming languages
are associated with more or fewer faults.
The results look at different scenarios and outline
how further custom analyses could be built atop the same model.
Finally, \autoref{sec:relwork} 
discusses related work 
and \autoref{sec:conclusions} concludes with a brief summary and closing discussion.

\section{Bayesian data analysis guidelines}
\label{sec:guidelines}

In the last decade, 
powerful Bayesian statistical analysis tools and languages %
have become
widely available together with computational resources adequate to run them~\cite{carpenter2017stan,ge2018turing,plummer2003jags}.
More recently, statisticians
have also been introducing and refining guidelines
on how to use these tools in a systematic way
to perform principled Bayesian data modeling~\cite{aczel20guidelines,schadBV20workflow,gabry2019visualization,gelman20workflow}.
In this section, we summarize these state-of-the-art guidelines %
while recasting them in a form suitable for empirical software engineering research.

A Bayesian model defines a statistical data-generating process
in terms of a \emph{prior} distribution of parameters $\theta$
and a \emph{likelihood} that certain data is observed for each value of the parameters.
\emph{Fitting} such a model on some empirical data $D$
then gives a \emph{posterior} distribution of the same parameters
that follows Bayes' theorem:
\begin{equation}
  \underbrace{P(\theta \mid D)}_{\text{posterior}} \quad \propto \quad \underbrace{P(D \mid \theta)}_{\text{likelihood}}\ \times\ \underbrace{P(\theta)}_{\text{prior}} \,.
  \label{eq:bayes}
\end{equation}
The posterior can then be used to compute the probability of other observations of interest in a predictive fashion.
Our previous work~\cite{FFT-TSE19-Bayes2} presented more details
about Bayes' theorem and the roles of prior, likelihood, and posterior.
In this paper, we focus on how to build a Bayesian model in practice:
Bayesian modeling involves choosing components in a way that is sound and principled, and that works for the data and domain that we are targeting.

\autoref{fig:workflow} illustrates a key idea of the \emph{guidelines} for Bayesian analysis presented here: developing a statistical model is a process of \emph{iterative refinement}, 
which starts from a very simple (possibly simplistic) \textit{initial model} that is gradually refined. 
Each iteration goes through a series of \emph{steps}
that assess the model's suitability in terms of the following characteristics:

\begin{figure}
    \centering
    \begin{tikzpicture}[
    bdastep/.style={rectangle,minimum height=10mm,text width=18mm,
    thick,font=\scshape\footnotesize,text=black,draw=black},
    anstep/.style={rectangle,rounded corners=2mm,,minimum height=10mm,text width=18mm,thick,font=\scshape\footnotesize,text=black,draw=black},
    node distance=7mm and 8mm,
    align=center]
    \node[align=center] (initial) {initial\\model};
    \node[bdastep,below=of initial] (plausible) {plausible?};
    \node[bdastep,right=of plausible] (workable) {workable?};
    \node[bdastep,right=of workable] (adequate) {adequate?};
    \node[anstep,below=of workable] (refine) {refine};
    \node[anstep,right=of adequate] (analyze) {analyze};
    \node[align=center,above=of analyze] (final) {analysis\\results};
    
    \begin{scope}[-latex,very thick]
    \draw (initial) -- (plausible);
    \draw (analyze) -- (final);
    \draw (plausible) -- node [above] {\checkOK} (workable);
    \draw (workable) -- node [above] (mp) {\checkOK} (adequate);
    \draw (adequate) -- node [above] {\checkOK} (analyze);
    \draw (workable) -- node [left] {\checkFail} (refine);
    \draw (adequate) -- node [below right=-1mm and 0mm] {\checkFail} (refine);
    \draw (plausible) -- node [below left=-1mm and 0mm] {\checkFail} (refine);
    \draw (refine.west) -| node [align=center,left] {refined\\ model}  (plausible);
    \end{scope}

    \node[anstep,above=of mp] (extend) {extend};
    \node[anstep,below=of analyze] (compare) {compare};
    \begin{scope}[very thick,every edge/.append style={-latex}]
       \begin{scope}[dashed]
       \draw (analyze) edge[bend right] (extend);
       \draw (extend) edge[bend right] node[align=center,above=4pt] {extended\\model} (plausible);
       \end{scope}
    \draw (compare) edge node [right] {``best''} (analyze);
    \draw (adequate.south east) edge node {\checkOK} ($(compare.north west)+(2pt,-2pt)$);
    \draw ($(adequate.south east)+(0pt,6pt)$) edge[bend left] node {\checkOK} ($(compare.north west)+(6pt,0pt)$);
    \draw ($(adequate.south east)+(-6pt,0pt)$) edge[bend right] node {\checkOK} ($(compare.north west)+(0pt,-6pt)$);
    \end{scope}
    \end{tikzpicture}
    \Description{Diagram with states \textsc{plausible?}, \textsc{workable?}, and \textsc{adequate?} (corresponding to the three main steps of the workflow), connected to states \textsc{refine}, \textsc{analyze}, and \textsc{compare}, which are the three main actions that one can perform on a model. The overall input is the ``initial model'' and the overall output are the ``analysis results''.}
    \caption{Process for Bayesian data analysis: starting from an initial model, 
    assess whether it is \emph{plausible}, \emph{workable}, and \emph{adequate}.
    If it lacks any of these characteristics, \emph{refine} the model by adding
    detail and features. Models that pass all checks can be fitted and
    used to answer the analysis's specific questions. 
    Different models that pass all checks can be rigorously \emph{compared}
    to select those that perform ``best'' according to suitable criteria.
    The outer loop (dashed arrows) indicates that an analysis's results 
    may also suggest to \emph{extend} an adequate model so that it
    can answer more precise, or just different, questions; this outer loop 
    is another source of multiple models that can be compared.}
    \label{fig:workflow}
\end{figure}
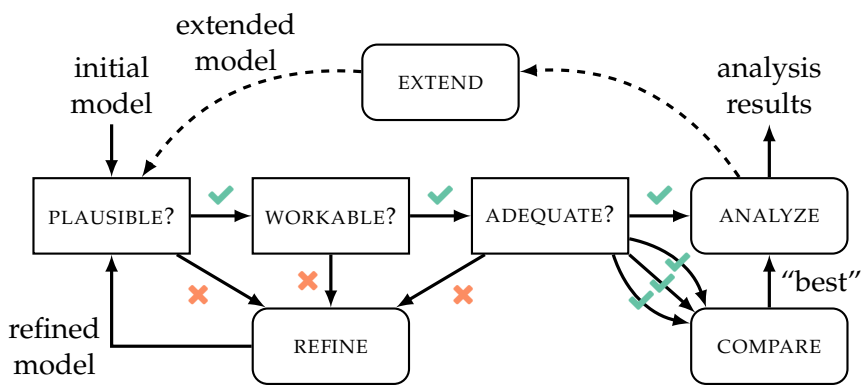

\begin{description}[labelwidth=15mm]
\item[Plausibility:] Is the model consistent with (expert) knowledge 
                  about the data domain?
                  
\item[Workability:] Can the model effectively 
    and accurately be fitted 
    using the available numerical algorithms?

\item[Adequacy:] Can the model capture the characteristics of
                 the empirical data?
\end{description}

These steps help assess the \emph{utility} (or suitability) of a model 
and its trade-offs. As we illustrate in Tables~\ref{tab:what-used} and~\ref{tab:artifacts},
each step puts additional requirements on a model,
and checks whether the model is well-equipped to faithfully capture the 
observed data and to analyze it. 
\autoref{tab:what-used} shows how 
each step broadens the scope of what model components are checked;
in particular, the actual empirical data is only used in the \emph{adequate} step,
whereas the previous steps generate simulated data using priors and likelihood.
\autoref{tab:artifacts} details how 
each step has a possible \emph{outcome} (what it establishes about the model),
which is supported by analysis \emph{artifacts} that document the step.
The following sections describe the steps, artifacts, and outcomes in some detail.
\autoref{sec:bda-pl} will apply the steps on the main case study
of programming language data.

\subsection{Modeling}
To make the description concrete, and thus easier to follow,
we illustrate what the steps compute on a toy problem:
predicting an adult person's height $h$ in centimeters.
To this end, we build this statistical model:
\begin{align}
\label{toy:likelihood}
   h &\sim\ \dist{Normal}(\mu, \sigma) \\
\label{toy:prior-mu}
   \mu &\sim\ \dist{Normal}(170, 50) \\
\label{toy:prior-sigma}
   \sigma &\sim\ \dist{HalfCauchy}(0, 1)
\end{align}
The following paragraphs introduce its components one by one.

\paragraph{Parameters.}
A Bayesian statistical model consists of three components:
parameters to estimate, likelihood, and priors.
In our example, the model's \emph{parameters} $\theta$
are the mean $\mu$ and standard deviation $\sigma$
of a person's height.
The data $D$ records the value of outcome variable $h$ for several persons,
which we can use to estimate $\mu$ and $\sigma$.
There are no predictor variables in this simplistic model, but otherwise these would 
also be recorded in the data for every person.

\paragraph{Likelihood.}
The \emph{likelihood} is a probability distribution
of the data $h$ given parameters $\mu$ and $\sigma$.
The simplest (yet extremely common)
choice is a \emph{normal} distribution,
which encodes no additional information about the data %
other than that it has a mean $\mu$ and a standard deviation $\sigma$.
This likelihood is defined by \eqref{toy:likelihood},
which is in fact a probability distribution of $h$ given $\mu$ and $\sigma$.

\paragraph{Priors.}
Finally, we need \emph{priors} for $\mu$ and $\sigma$.
Specifying a prior means defining an initial probability distribution
for a parameter of the model---a probability distribution
without any dependency on the data. %

The least informative priors are completely \emph{flat} distributions,
which assign the same infinitesimal probability to any value of the parameter;
this is the default behavior in frequentist statistics.
A flat prior for $\mu$, for example, would be a uniform distribution with support from $-\infty$ to $+\infty$.
Flat priors are usually a poor choice:
first, since they stretch a probability distribution over an infinitely large support, they tend to generate infinitesimal probabilities that 
may cause \emph{numerical} rounding errors;
second, a prior with no information whatsoever about the realistic
parameter domain is prone to \emph{overfitting} the data.
We can see it clearly even in the simple example of estimating heights:
a flat prior would give the same a priori probability to
height values $-10$, $170$, and $10^{9}$,
but only the second value is a plausible human height!

A better choice are \emph{weakly informative} 
priors, which still carry very little specific information 
but perform much better than flat priors computationally
and protect against overfitting the data.
As prior for $\mu$, we select the normal distribution \eqref{toy:prior-mu}, 
with mean $170$
and standard deviation $50$;
this means that we expect most heights to be between
$20 = 170 - 3\cdot 50$
and $320 = 170 + 3\cdot 50$
centimeters.
This is still an extremely broad range of values, but
it favors values that are in the ballpark of realistic human heights.
By the way, there is nothing special about the values $170$ and $50$:
the priors are only a starting point, which should just
identify a plausible range of heights without being unnecessarily constraining.
Different, reasonable choices for the priors would still lead to very similar outcomes.

A prior for $\sigma$ should rule out negative values
(that is, assign zero probability to them), 
since a standard deviation must be a nonnegative number.
A common choice for priors of standard deviations
is a so-called half-Cauchy distribution,
which is a truncated Cauchy.
Precisely, we set the first (location) parameter of the prior \eqref{toy:prior-sigma}
for $\sigma$ to zero,
so that the distribution's support is restricted to the nonnegative reals.
We set the second (scale) parameter to one, which spreads out
the probabilities smoothly while still preferring moderate values of $\sigma$.

Bayesian data analysis tools can often suggest
\emph{default} weakly informative priors that may work well in many cases.
In the analysis of \autoref{sec:bda-pl}, we will define our priors---following standard recommendations~\cite{mcelreath2020statistical}---but
very often using default priors would have lead to overall similar results.
In any case, the plausibility checks described next will validate our choice of priors.

\begin{table}
    \centering
    \setlength{\tabcolsep}{4pt}
    \begin{tabular}{l cc cc c}
    \toprule
    \multirow{2}{*}{\textsc{step}} & \multirow{2}{*}{\textsc{prior}} & \multirow{2}{*}{\textsc{likelihood}} & \textsc{alternative} & \textsc{empirical} 
    & \textsc{new data} \\[-3pt] 
    &&& \textsc{models} & \textsc{data} & \textsc{for prediction} \\
    \midrule
    \textbf{plausible?} & \checkmark & & \\
    \textbf{workable?} & \checkmark & \checkmark & \\
    \textbf{adequate?} & \checkmark & \checkmark && \checkmark \\
    \textbf{compare} & & & \checkmark & & \\
    \cmidrule(lr){1-1}
    \textbf{analyze} & \checkmark & \checkmark && \checkmark & \checkmark \\
    \bottomrule
    \end{tabular}
    \caption{For each step of a Bayesian statistical analysis 
    (checks of plausibility, workability, adequacy, model comparison, and analysis),
    the model components (\textsc{prior} and \textsc{likelihood}),
    competing \textsc{alternative} models, 
    and kinds of \textsc{data} (\textsc{empirical} and \textsc{new for prediction})
    that the step primarily \emph{tests}.}
    \label{tab:what-used}
\end{table}

\begin{table}
\centering
\footnotesize
\renewcommand{\arraystretch}{1.3}
\begin{tabularx}{\textwidth}{lXX}
    \toprule
    \textsc{step} & \multicolumn{1}{c}{\textsc{artifacts}} & \multicolumn{1}{c}{\textsc{outcome}} \\
    \midrule
    \textbf{plausible?} &
        \begin{enumerate*}
        \item Prior predictive simulation plots; 
        \item Justification for priors if they disallow certain values.
        \end{enumerate*}
        &
        The priors allow a broad range of possible values and
        give low probability to values that are unlikely to occur in the domain.
        \\
    \textbf{workable?} &
        \begin{enumerate*}
        \item Simulation-based calibration of $z$ score and shrinkage;
        \item Fitting diagnostic metrics.
        \end{enumerate*}
        &
        Fitting the model works computationally and does not
        exhibit pathological behavior.
        \\    
    \textbf{adequate?} &
        Posterior predictive checks plots.
        &
        The model can generate data similar to the empirical observations.
        \\    
    \textbf{compare} &
        Information-criteria ranking and scores of competing alternative  models.
        &
        The chosen model achieves a bias-variance trade-off better than the alternative 
        models.
        \\    
    \cmidrule(lr){1-1}
    \textbf{analyze} & 
        \begin{enumerate*}
        \item Posterior plots based on the empirical data;
        \item Distribution plots and summary statistics of
              any domain-specific variables of interest.
        \end{enumerate*}
        &
        Quantitative answers to the analysis's specific questions.
    \\
    \bottomrule
\end{tabularx}
\caption{The \textsc{artifacts} that are typically produced,
and the \textsc{outcome} that follows from each step of a Bayesian statistical analysis (when the step succeeds).}
\label{tab:artifacts}
\end{table}

\subsection{Plausible model}
\label{sec:plausible-def}

Once we have chosen parameters, likelihood, and priors 
our model definition includes all required parts.
Then, we can ``run'' the model---that is, sample from it---and 
analyze how likelihood, data, and priors constrain the model parameters of interest~\cite{gelmanSB17prior}. 
The first step of this analysis focuses on 
the priors, which should be neither too constraining nor unreasonably permissive.
This step is typically called \emph{prior predictive simulations} or \emph{prior predictive checks}~\cite{schadBV20workflow,mcelreath2020statistical}
and works as follows:
sample the priors broadly;
using the sampled distribution 
(ignoring the actual empirical data, which are not used in this step),
run the model to get a distribution of the model variables of interests
(typically, the outcome variable);
check that this distribution is plausible for the variables that it
measures.

The key principle %
is that, if the priors are properly chosen,
this process should determine a distribution that allows
all plausible values for the variables
but gives vanishing small probability to values that 
are practically impossible or contradict established scientific knowledge.
In summary, prior predictive checks answer the question: \textsl{Does sampling from the priors lead to a plausible range of parameter values?}

Here is how prior predictive simulation would work on our toy example.
First, we sample random values for parameters $\mu$ and $\sigma$ from their
prior distributions \eqref{toy:prior-mu},\eqref{toy:prior-sigma}.
We then plug each sampled pair of values $\overline{\mu}, \overline{\sigma}$
into the likelihood \eqref{toy:likelihood}
and sample values of $h$ from $\dist{Normal}(\overline{\mu}, \overline{\sigma})$.
Since the outcome variable $h$ measures an adult person's height, 
the priors should be such that this sampled distribution of $h$ freely allows 
heights between, say, 0 and 300 cm, whereas it disallows negative heights and assigns very small probabilities to heights above 300 cm---since no human on record
has ever been that tall.

Prior predictive simulation is not cheating~\cite{mcelreath2020statistical}. As long as we do not set priors based on the actual empirical data that we are going to analyze, but only through what we know about the data \emph{domain} independent of how we measured it, it is sensible to use our existing knowledge to rule out priors that would lead to impossible or clearly implausible results. %
Seen in this light, the possibility of choosing priors is a big advantage of Bayesian analysis
that is highly valuable for any empirical science.
Using existing knowledge to guide new analyses,
we can develop sequences of studies that, taken together, 
progressively sharpen knowledge in a specific area.
The alternative is that every software engineering research contribution
remains an ``island unto itself''
without clear connections to the related literature and the field as a whole~\cite{FFT-TSE19-Bayes2}.

The reasons for choosing certain priors that make the model \emph{plausible} should be explicitly justified.
In practice, and to the extent that it is possible, empirical software engineering studies should explicitly state which published results, common sense, or ``folk knowledge'' justify the choice of priors and their plausibility behavior. 
\autoref{sec:plausible} demonstrates how to do that for the paper's case study.

Selecting informative priors gives Bayesian statistics more flexibility,
but does not limit its applicability.
When very little is known about the problem domain---for example,
in an exploratory first study about a certain practice---one
can always fall back to using completely uninformative priors, which
require no specific knowledge, and hence are vacuously plausible.
When prior knowledge exists, however, defining more selective priors
can help sharpen the model and specialize it to the characteristics
of the analysis domain.

\subsection{Workable model}
\label{sec:workable-def}

Once we have ascertained that the chosen 
priors are consistent with plausible parameter values,
the second step checks whether our model \textit{works} com\-pu\-tat\-ion\-al\-ly---that is,
\emph{fitting} the model does not incur divergence or other numerical problems, and 
the fitting process eventually reaches a stationary state 
that properly identifies a posterior distribution.

An emerging technique to do so is 
\emph{simulation-based calibration}~\cite{talts2018validating,schadBV20workflow},
which relies on a consistency property of Bayesian models:
first, simulate parameter and data values 
from the priors and likelihood as done in prior predictive simulations;
then, using Bayes' theorem, 
combine the simulated parameter and data samples 
to get a \emph{posterior} distribution of the model's parameter;
if the model is consistent, 
the posterior distribution obtained in this way should resemble the prior distribution.
To perform simulation-based calibration on our toy example,
we would
\begin{enumerate*}[label=(\emph{\roman*})]
\item sample parameter values from the priors;
\item use those to build samples of the outcome variable $h$;
\item use these outcome samples as data (instead of the actual empirical data) 
and combine them again with priors and likelihood
using Bayes' theorem \eqref{eq:bayes}.
\end{enumerate*}
These steps give a new sampled distribution of the parameters $\mu$ and $\sigma$,
which we compare with that obtained by sampling the priors directly in the first step.

While promising, simulation-based calibration
is a cutting-edge technique that is still
undergoing major developments;
none of the statistical analysis tools 
that are more widely used for Bayesian analysis
support it out-of-the-box.
Instead, these tools offer other metrics to assess workability that
are specific to the fitting algorithms based on dynamic Hamiltonian Monte Carlo which they implement.
Here are the metrics that are usually available, and how they help
us assess workability:

\begin{itemize}
\item A \emph{divergent transition} in the sequences of samples indicates
a possible numerical error; workable models should have few divergent transitions---ideally none. %

\item The sampling process is repeated a few (usually 2--4) times independently; each sequence of sampling is called a \emph{chain}.
In a workable model, different chains should be statistically similar:
the ratio $\widehat{R}$ of within-to-between chain variance
should converge to 1 as the number of samples grows.
A common rule of thumb for finite sampling is that $\widehat{R}<1.01$, which indicates a stationary posterior distribution.

\item The \emph{effective sample size} is the fraction of 
all samples that are independent, that is not autocorrelated.
We typically want it to be at least 10\% for each parameter we estimate 
(and that the absolute number of independent samples be a few hundreds); lower values may indicate that sampling is ineffective.

\item Finally, we can also visually inspect the plots that trace the 
samples in every chain. When the different lines look mixed up (like a ``hairy caterpillar''~\cite{mcelreath2020statistical}),
it is one more sign that the fitting process works well.
\end{itemize}
\autoref{sec:workable} uses these metrics 
to analyze the workability of our models.

When a workability check fails, it suggests that there is a mismatch
between the model and the algorithm used to fit it.
Sometimes, this is due to the data---for instance it is too sparse to effectively sample from it.
More commonly, it indicates that the model itself is unsuitable for the analysis at hand.
A clear example is the problem of \emph{multicollinearity}:
when two variables are strongly correlated, their exact contribution to the outcome is undetermined;
thus, the model may not be workable because it cannot be used to discover a definite
value for each variable independent of the other.

\subsection{Adequate model}
\label{sec:adequate-def}
If the previous analysis steps were successful, 
we determined that the priors are sensible (\emph{plausibility})
and that fitting the model is a converging process (\emph{workability});
it remains to check whether the model adequately captures reality.
The third step thus fits the model using the actual empirical data 
(which was not used in the previous two steps)
and performs \emph{posterior predictive checks}:
using the posterior distribution of parameters fitted on the actual empirical data, %
simulate new observations and compare them to the data.
If the two are consistent, it means that the model can generate data
similar to the observed data, and hence it captures the empirical observations \emph{adequately}.

Here is how posterior predictive checks would work on our toy example.
Similarly as in simulation-based calibration, we combine data and prior samples using Bayes' theorem \eqref{eq:bayes};
the key difference is that we now use the actual observed data 
(the height of real people) instead of simulated data.
This gives a posterior predictive distribution
of parameters $\mu$ and $\sigma$,
which, in turn, we sample; 
then, we plug the sampled parameter values 
into the likelihood
to get a distribution of $h$---the so-called
\emph{posterior predictive distribution}, since it expresses
the information about the posterior indirectly in terms of prediction
of model (outcome) variables.
In an adequate model, the posterior predictive distribution 
generates data somewhat similar to the actual observed data.

\autoref{sec:adequate} discusses the results of posterior predictive checks
on the programming language case study.

\subsection{Model comparison}
\label{sec:comparison-def}

Information criteria such as %
WAIC (Widely Applicable Information Criterion, also known as Watanabe-Akaike Information Criterion)~\cite{watanabe10waic} and PSIS-LOO (Pareto-Smoothed Importance Sampling Leave One Out validation)~\cite{vehtariGG17loo} 
assess a kind of relative adequacy %
by measuring deviance or other information-theoretic metrics
between a model's predictions and the data.
In a nutshell, these metrics assess how well each model
performs out-of-sample predictions compared to other competing models.
Thus, information criteria measures 
are \emph{relative}:
they are useful to compare the adequacy of a model relative to another but cannot
gauge a model's adequacy in absolute terms.
\autoref{sec:compare} uses information criteria to compare different models
for the programming language data analysis.

\subsection{Iterative refinement} 
\label{sec:iteration-def}
After a candidate model goes through the steps described above,
we have a clear understanding of its strengths and weaknesses.
When the model fails specific steps, 
we also learn what aspects we have to change to refine it:
the priors of an implausible model need changing;
an unworkable model needs to be refactored in a way that works computationally;
an inadequate model may require more information (typically
in the form of additional variables or parameters) for it to be consistent with the data (for 
example, to properly capture inter-group variability).

Model design is an \emph{iterative} process
which gradually refines an initial model to improve it.
Usually, we start from a deliberately 
very simple model and make it more complex as needed~\cite{schadBV20workflow}.
However, we can also do the opposite: 
start from a so-called \emph{maximal} model,
and then simplify it 
as long as it retains the characteristics of plausibility, workability, and adequacy~\cite{piironenPV2020projpred}.
In practice, we may even alternate simplification and refinement (detail-adding) steps
starting from a canonical model~\citep[pp.\ 103--104]{neal96bayes} until we are satisfied with the results.

The presentation of the results of a Bayesian data analysis
need not discuss the models in the same order in which they were designed and evaluated;
it does not even need to present all models,
but can simply present the final model as long as its choice can 
be soundly justified a posteriori (and, preferably, a reproducibility package exists).
Regardless of how we choose to present the overall outcome of an analysis, %
considering different models
expands the flexibility of the modeling process,
supports making informed choices about each aspect of a model,
and helps focus on and quantify the relative benefits of
each model in terms of the trade-offs that matter for the ongoing analysis.

\subsubsection{Uniqueness and optimality of models}
\label{sec:uniqueness-optimality}
When should we stop refining our model?
Paraphrasing George Box's famous aphorism~\cite{allModelsWrong-1,allModelsWrong-2},
we could say that the goal of statistical modeling is building a useful model,
not a correct one.
In other words, we cannot expect that
following our guidelines leads to designing a unique or optimal model.

Model comparison can identify which models perform better predictions than other models,
but it cannot assess a model's absolute predictive capabilities.
The steps in \autoref{fig:workflow}
make up a \emph{validation} process,
which can identify a model's shortcomings or confirm that it is of suitable quality;
they cannot say anything about the infinitely many \emph{other} models
that were not considered.
Building useful models still requires human intuition, knowledge, and ingenuity---skills
that no supporting process can completely replace.

\subsection{Tools for Bayesian data analysis}
\label{sec:tools}

Let us briefly mention which tools are available to support the kind of Bayesian data analysis process
that we discuss in this paper.
Stan~\cite{carpenter2017stan} and JAGS~\cite{JAGS}
are state-of-the-art frameworks that offer a probabilistic language to express Bayesian models
and implement very efficient algorithms to fit such models on data.
Commonly, one uses these frameworks through a front-end library in a high-level programming language suitable for data analysis.
Libraries such as \texttt{brms} for R~\cite{brms}, \texttt{Stan.jl} for Julia~\cite{Stan-jl}, and PyStan for Python~\cite{PyStan}
provide a rich interface to Stan, including support for the main steps of our guidelines (for example, prior predictive simulations).
\texttt{Turing.jl} for Julia~\cite{Turing-jl}
also provides a high-level interface to perform Bayesian data analysis,
but includes its own implementation of Bayesian sampling instead of relying on Stan's
(which may offer some advantages in terms of flexibility and generality for the most advanced applications).
\section{Bayesian data analysis of programming language data}
\label{sec:bda-pl}

Equipped with a high-level understanding of the modeling guidelines 
that we outlined in \autoref{sec:guidelines}, 
we apply them to perform the analysis of the FSE data.
The overall outcome of the work described in this section will be a 
carefully designed, suitable statistical model of this data.
In \autoref{sec:bda-results}, we will analyze this model 
to understand what it tells us about the original questions 
on programming languages and code quality.

To mitigate the risk of mono-operational bias,
the first author prepared the data for analysis in R, 
and the second author developed the first complete analysis,
which then the first and third author revised.
Finally, all three authors validated
the final revised analysis, which is presented here.
In addition, the second author did not
read the publication that originated the dataset~\cite{FSE}
or its reanalysis~\cite{TOPLAS}
until after completing the first complete analysis.
This reduced the chance that the others'
design decisions, or some characteristics of the data
they highlighted, biased our application 
of the modeling guidelines.\footnote{%
We used Stan through its \texttt{brms} R front-end to
perform the analysis described in the rest of the paper.
}

\subsection{Data}
\label{sec:data}
FSE's authors released the original dataset---obtained by mining information from GitHub repositories---upon request from TOPLAS's authors. 
TOPLAS performed first a repetition of FSE's analysis on the same dataset,
and then a reanalysis on a revised dataset obtained by ``alternative 
data processing and statistical analysis to address what [they] identified
as methodological weaknesses of the original work''~\cite[Sec.~4]{TOPLAS}.
The main difference between FSE's original dataset and TOPLAS's revised dataset
is that the latter removes some duplicated data, 
TypeScript projects (which often do not include much actual TypeScript code),
and the V8 project (whose JavaScript code in the dataset is mostly tests).
Finally, TOPLAS's replication package includes FSE's original dataset
alongside TOPLAS's revised dataset. %

In our analysis, we focus on the original FSE dataset,\footnote{Which we obtained from TOPLAS's public replication package (available at  \url{https://github.com/PRL-PRG/TOPLAS19_Artifact}).} because we
would like to see whether a Bayesian data analysis can help spot
issues and inconsistencies in the data 
that %
may hinder replication attempts---and, conversely, that
may make replication run-of-the-mill if addressed early on.
Our replication package includes all analysis details, 
including the results of fitting the same models on TOPLAS's revised dataset
(which we do not discuss here for brevity).

FSE's dataset includes information 
about \numprint{1578165} commits,
which we group by project and language
giving \numprint{1127} datapoints.
The attributes that are relevant for our analysis are:

\begin{description}[labelwidth=18mm]
\item[\var{project}:] the project's name
\item[\var{language}:] the used programming language
\item[\var{commits}:] the total number of commits in the project
\item[\var{insertions}:] the total number of inserted lines in all commits
\item[\var{age}:] the time passed since the oldest recorded commit in the project
\item[\var{devs}:] the total number of users committing code to the project
\item[\var{bugs}:] the number of commits classified as ``bugs''
\end{description}

The values of attributes \var{commits}, \var{insertions}, \var{age}, and \var{devs}
vary greatly between projects.
When this happens, it is customary to transform the data using a logarithmic function,
so that the variability is over a smaller range whose unit corresponds to an order of magnitude.
Both FSE's and TOPLAS's analyses log-transformed these attributes;
we do the same: henceforth, 
\var{commits}, \var{insertions}, \var{age}, and \var{devs}
represent the natural \emph{logarithm} of the total number of commits, inserted lines, and so on.

\begin{figure}
  \centering \includegraphics[width=\textwidth]{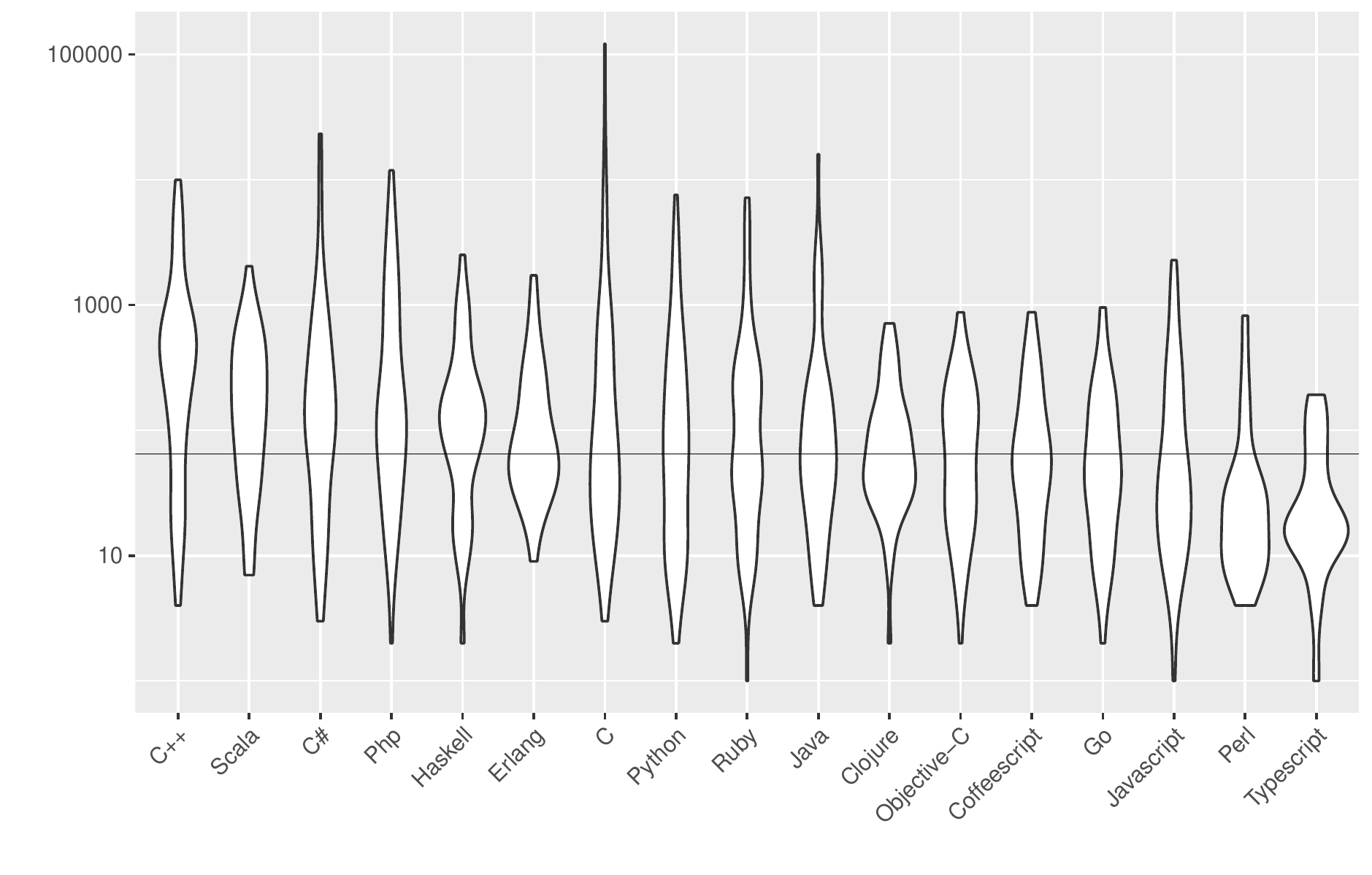}
  \Description{For each programming language, the figure shows a
    violin plot of the distribution of bugs in projects written in
    that language.  The plots are sorted in descending order, from the
    programming language with the largest median bugs per project
    to that with the smallest: C++ (largest), Scala, C\#, Php, Haskell, Erlang, C, Python, Ruby,
    Java, Clojure, Objective-C, CoffeeScript, Go, JavaScript, Perl,
    and TypeScript (smallest).
    The figure also shows the median number of bugs per project in all dataset,
    which is around 480.}
  \caption{Violin plots of the distributions of number of bugs per
    project for each programming language in the original FSE dataset.  Languages are
    sorted, left-to-right, by decreasing values of the distributions'
    medians.  The vertical axis's scale is logarithmic in base $10$.
    An horizontal line marks the median number of bugs per project
    across all languages.}
    \label{fig:raw-data}
\end{figure}

\autoref{fig:raw-data} provides an overview of the FSE dataset,
showing the distribution of bugs per project grouped by programming language.
Visualizing the raw data can be useful to get a broad idea of what is in the dataset;
however, the information that such visualizations provide is mostly qualitative
and we should be aware of its limitations.
If one includes all data, any outliers may skew the picture at extreme values;
conversely, if one excludes some data,
deciding which data to exclude is itself a source of possible bias,
and discards potentially useful information thus increasing uncertainty.
In this dataset specifically, projects vary broadly
in terms of size and other characteristics.
\autoref{fig:raw-data} conflates these differences,
and hence a comparison of different languages based on it may be misleading.
We could display a subset of the data that only includes projects with homogeneous characteristics;
however,
doing so would drop significant amounts of information,
introduce a somewhat arbitrary partitioning (what projects are ``similar''?), 
and increase the risk of overfitting other accidental characteristics of the data.
By abstracting the information in the raw data and combining it with expert knowledge,
a suitable statistical model can lessen several of these problems,
thus supporting more robust and general inferences about the impact of programming languages.

\subsection{Modeling}
\label{sec:modeling}

We build three models---$\model_1$, $\model_2$, and $\model_3$---of increasing complexity.\footnote{As discussed in \autoref{sec:iteration-def}, for clarity 
we present the three models at once but we actually 
designed them over several iterated applications of the guidelines.} %
\autoref{tab:steps-pl} summarizes the outcome of the steps in \autoref{fig:workflow} for the three models.
Mirroring \autoref{tab:artifacts}'s structure,
\autoref{tab:artifacts-pl} outlines the artifacts produced in each step,
and the conclusions that the analysis draws about each model's suitability.
The rest of this section details the models and the outcome of the guidelines' 
suitability checks presented in \autoref{sec:guidelines}.
This section presents the models and the outcome of their suitability analysis in detail;
later, \autoref{sec:modeling-2} will discuss, at a higher level, what this analysis reveals about the relations between model features and data.

As we remarked in \autoref{sec:scope},
our guidelines can be used to validate different kinds of models.
We consider these three models because they belong to a widely used family of statistical models,
and for their similarity with the models of FSE and TOPLAS.
An analysis with different goals or done by analysts with different expertise %
could end up building very different kinds of models---but they should still undergo the same validation steps.

\begin{table}
    \centering
    \begin{tabular}{l ccc}
    \toprule
    \textsc{step} & $\model_1$ & $\model_2$ & $\model_3$ \\
    \midrule
    \textbf{plausible?} & \checkOK & \checkOK & \checkOK \\
    \textbf{workable?}  & \checkOK & \checkOK & \checkOK \\
    \textbf{adequate?}  & \checkFail & \checkOK & \checkOK \\
    \textbf{compare} &  -- & \checkFail & \checkOK \\
    \bottomrule
    \end{tabular}
    \caption{All three models are \emph{plausible} and \emph{workable},
    but model $\model_1$ is \emph{not adequate} because it cannot accurately capture the regular features of the dataset. Model \emph{comparison} between $\model_2$ and $\model_3$ shows that the latter performs much better concerning out-of-sample predictions, and hence
    we will use $\model_3$ for the rest of the analysis.}
    \label{tab:steps-pl}
\end{table}

\begin{table}
\centering
\footnotesize
\renewcommand{\arraystretch}{1.3}
\begin{tabularx}{\textwidth}{lXX}
    \toprule
    \textsc{step} & \multicolumn{1}{c}{\textsc{artifacts}} & \multicolumn{1}{c}{\textsc{outcome}} \\
    \midrule
    \textbf{plausible?} &
        \begin{enumerate*}
        \item Prior predictive simulation plots in \autoref{fig:ppc}; 
        \item Justification for priors: typical relations between project size and
        number of known bugs~\cite{scholzT20LOCs}.
        \end{enumerate*}
        &
        The priors of all three models 
        allow a very broad range of possible values for the number of bugs that
        may exist; extremely high numbers are still possible but with low
        probability.
        \\
    \textbf{workable?} &
        Fitting diagnostic metrics reported in~\autoref{sec:workable}: $\widehat{R}$, effective sample size, 
        no divergent transitions, trace plots (details in the replication package).
        &
        Fitting all three models works computationally and reaches convergence.
        \\    
    \textbf{adequate?} &
        Posterior predictive checks plots in \autoref{fig:postpc}. 
        &
        Model $\model_1$ cannot 
        generate a distribution similar to
        that observed in the data (top plot in \autoref{fig:postpc}),
        whereas models $\model_2$ and $\model_3$ can.
        \\
    \textbf{compare} &        
        Information-criteria scores of competing models in \autoref{tab:loo-differences}.
        &
        Model $\model_3$ clearly outperforms
        $\model_2$ in how it can predict data out of the sample used for fitting.
        \\    
    \cmidrule(lr){1-1}
    \textbf{analyze} & 
        \begin{enumerate*}
        \item Posterior plots based on the empirical data in Figures~\ref{fig:ranks-max-median-min} and \ref{fig:effect-sizes};
        \item Distribution plots and summary statistics of
              domain-specific variables of interest in
              Figures~\ref{fig:insertions-effects} and \ref{fig:python-ruby-comparison}.
        \end{enumerate*}
        &
        Quantitative answers to the analysis's specific questions in \autoref{sec:bda-results}.
    \\
    \bottomrule
\end{tabularx}
\caption{A summary of the \textsc{artifacts} produced by the analysis
of the three models,
and the \textsc{outcome} of each step of the analysis 
in terms model suitability. This summary instantiates \autoref{tab:artifacts}
for the programming language data analysis.}
\label{tab:artifacts-pl}
\end{table}

\subsubsection{Likelihood (and parameters)}
\label{sec:models-likelihood}

Generalized linear models are a broad category of statistical models
that are so flexible that they can be ``applied to just about any problem''~\cite{GelmanHill07-MultilevelModels}
when modeling empirical data.
The likelihood of a generalized linear model is %
a probability distribution over certain parameters, 
which are generalized linear functions of the variables chosen as predictors.
The values drawn from the distribution correspond to the outcome that we 
are modeling.

\paragraph{Distribution family.}
In our case, the \emph{outcome} variable is \var{bugs}, which 
always is a nonnegative integer.
Therefore, we should select a likelihood distribution
suitable for ``count\-ing''---that is, one in the Poisson family. 
The single-parameter Poisson
is the distribution in this family with the highest information 
entropy~\cite{jaynes03prob}, and hence it should be the customary initial choice. %

Nevertheless, building a model using the single-parameter Poisson
quickly reveals that it cannot
account for the fact that 
the distribution of \var{bugs} in the data is \emph{overdispersed}: its 
mean $\mu_\var{bugs} = 501$ is much smaller 
than its variance $\sigma^2_\var{bugs} = \numprint{15031006}$.
This justifies 
selecting the slightly more complex \emph{negative binomial} distribution $\dist{NegativeBinomial}(\lambda, \phi)$.  %
The two parameters $\lambda$ and $\phi$ represent\footnote{\url{https://mc-stan.org/docs/2_20/functions-reference/nbalt.html}} 
rates that together determine
the distribution's
mean $\lambda$ and variance $\lambda + \lambda^2/\phi$, which can take different values to accurately capture overdispersion.
This is in contrast to the Poisson distribution whose mean and variance coincide.
The negative binomial distribution is also the same distribution selected, for the same
reason, by both FSE's original analysis and TOPLAS's reanalysis.

\begin{figure}
    \centering
    \small
    \begin{subfigure}[b]{0.35\textwidth}
    \centering
    \begin{align*}
    \var{bugs}_i &\sim\ \dist{NegativeBinomial}(\lambda_i, \phi) 
    \\
    \log(\lambda_i) &=\ \Pi_i + L_i \\
    \Pi_i &=\ \alpha \\
    L_i  &= \ \alpha_{\var{language}_i}
    \end{align*}
    \caption{Model $\model_1$}
    \label{fig:m1}
    \end{subfigure}
    \hspace{5mm}
    \begin{subfigure}[b]{0.35\textwidth}
    \centering
    \begin{align*}
    \var{bugs}_i &\sim\ \dist{NegativeBinomial}(\lambda_i, \phi) 
    \\
    \log(\lambda_i) &=\  \Pi_i + L_i
    \\
    \Pi_i &=\ \alpha + {\color{acol} \coef{commits} \cdot \var{commits}_i}
          + {\color{acol} \coef{insertions} \cdot \var{insertions}_i} \\
        &\qquad + {\color{acol} \coef{age} \cdot \var{age}_i}
        + {\color{acol} \coef{devs} \cdot \var{devs}_i}
    \\
    L_i  &= \ \alpha_{\var{language}_i}
    \end{align*}
    \caption{Model $\model_2$}
    \label{fig:m2}
    \end{subfigure}
    \begin{subfigure}[b]{0.35\textwidth}
    \centering
    \begin{align*}
    \var{bugs}_i &\sim\ \dist{NegativeBinomial}(\lambda_i, \phi) 
    \\
    \log(\lambda_i) &=\  \Pi_i + L_i + P_i
    \\
    \Pi_i &=\ \alpha + {\color{acol} \coef{commits} \cdot \var{commits}_i}
          + {\color{acol} \coef{insertions} \cdot \var{insertions}_i} \\
        &\qquad + {\color{acol} \coef{age} \cdot \var{age}_i}
        + {\color{acol} \coef{devs} \cdot \var{devs}_i}
    \\
    L_i &=\ \alpha_{\var{language}_i} + {\color{bcol} \coef{commits}_{\var{language}_i} \cdot \var{commits}_i} \\
          &\qquad + {\color{bcol} \coef{insertions}_{\var{language}_i} \cdot \var{insertions}_i}
          + {\color{bcol} \coef{age}_{\var{language}_i} \cdot \var{age}_i} \\
          &\qquad + {\color{bcol} \coef{devs}_{\var{language}_i} \cdot \var{devs}_i}
    \\
    P_i  &= \ {\color{bcol} \alpha_{\var{project}_i}}
    \end{align*}
    \caption{Model $\model_3$}
    \label{fig:m3}
  \end{subfigure}
  \Description{Equations describing the likelihoods of $\model_1$, $\model_2$, and $\model_3$ given as text.}
    \caption{The likelihoods of statistical models $\model_1$, $\model_2$, and $\model_3$.
    Colors highlight the terms that are added to each model compared to the previous ones.}
    \label{fig:models123}
\end{figure}

\paragraph{Model $\model_1$.}
The first model we consider, called $\model_1$, is very simple:
it assumes that the rate $\lambda$ 
is a function of two terms only.
The first term $\Pi$ is a constant intercept $\alpha$;
the symbol $\Pi$ highlights that it is a population-level term.
The second term $L$ is an additional intercept $\alpha_{\var{language}}$ 
that depends only on the \emph{language} used in each observation;
the symbol $L$ highlights that it is a language-level term.
\autoref{fig:m1} shows $\model_1$'s overall likelihood,  
where the logarithm function \emph{links}\footnote{%
In other words, the link function converts measures from the probability space to the outcome space.} the linear function of the parameters 
and $\lambda$
so that the latter is always a nonnegative number---as it should be in a ``counting'' distribution.

Model $\model_1$ is obviously too simple to capture the variability in the data
with high accuracy.
Nonetheless, it is a useful starting point to understand the key relations between variables
and to bootstrap the process that leads to incrementally more refined and precise models.
In its simplicity, it highlights that the key predictor (the ``treatment'') is
the programming language used in each project, whose relation with the number of bugs
we would like to capture.
Finally, even a simplistic model serves as a useful \emph{baseline} to compare
to more complex models---as a sanity check that the additional complexity that
we are going to add to the models brings measurable improvements over the baseline.

\paragraph{Model $\model_2$.}
The second model we consider, called $\model_2$, is a standard
linear-regressive model with negative binomial likelihood.
Model $\model_2$'s population-level term $\Pi$ 
is a linear function with intercept $\alpha$ and a slope $\beta$ for
each predictor variable \var{commits}, \var{insertions}, \var{age}, and \var{devs}.
In addition, like model $\model_1$, $\model_2$ includes a language-level term $L$
that consists of an intercept that depends on the language used in each observation 
(a so-called ``varying intercept'' model~\citep{gelman2020regression}).
\autoref{fig:m2} shows $\model_2$'s overall likelihood.

Model $\model_2$ is the closest to the
regressive models used in FSE and TOPLAS.
The only difference is how each model accounts 
for the dependence on the programming \emph{language}:
FSE and TOPLAS use different kinds of \emph{contrasts}~\citep{FSE, TOPLAS},
whereas we simply add an intercept language-level term---thus making our models \emph{multilevel}~\cite{GelmanHill07-MultilevelModels}.
Multilevel modeling comes natural with Bayesian statistics,
both because we do not have to worry too much about adding layers to the model
(unlike with frequentist techniques, changing such characteristics of the model does not
require changing the fitting algorithm)
and because we can just model the quantities of interest directly
and \emph{compute} any derived quantity 
\emph{after} we fit the model's \emph{posterior} distribution
(unlike with frequentist techniques, which mostly provide only point estimates without distributional information).

\paragraph{Model $\model_3$.}
The third model we consider, called $\model_3$, is a multilevel model
that tries to capture the effect of the programming language with greater detail.
Model $\model_3$'s population-level term $\Pi$ is identical to $\model_2$'s.
Its language-level term $L$ is considerably more complex, since
it introduces a linear model with different intercepts $\alpha_{\var{language}}$ 
and slopes $\beta_{\var{language}}$ for each programming \var{language} (a so-called
``varying intercepts and varying slopes'' model, also commonly known as ``varying effects'' model~\citep{gelman2020regression}).
Unlike the population-level term $\Pi$, 
the language-level term $L$ \emph{pools}
the information about each data cluster---where clusters are identified by the
used programming language.
Since it clusters by programming language, 
this partial pooling may help capture more accurately 
the effects of choosing a programming language instead of another;
at the same time,
it also shares information 
among clusters so that
some information from larger clusters (languages with many projects)
can sharpen the information from small\-er clusters (languages with fewer projects).  This also means that partial pooling helps protect from \emph{overfitting}, as learning takes place
first separately on each cluster, 
and then is ``regularized'' 
by sharing its results among different clusters.

The three models' focus on the programming language reflects
our intuitive expectation that the relation between programming languages
and proneness to bugs is an important one---regardless of 
whether it turns out to be significant or negligible in the end.
At the same time, adding predictors other than the programming language
\emph{accounts for} confounding factors that may have a stronger correlation with
the number of bugs.
But what if the intrinsic differences between \emph{projects} 
turn out to dominate the discrepancies in code quality?
For instance, 
different projects may have wildly different protocols to
report, triage, and fix bugs, which might have an effect on the observed 
number of bugs.

In order to hedge against this possible confounding factor,
model $\model_3$ also includes a term $P$: 
an additional intercept $\alpha_{\var{project}}$ 
that depends only on each observation's \emph{project};
the symbol $P$ highlights that it is a project-level term, which will help in quantifying the intrinsic variability across projects.
\autoref{fig:m3} shows $\model_3$'s overall likelihood.

\subsubsection{Priors}
\label{sec:models-priors}

As we demonstrated in the previous section,
choosing the likelihood typically requires making justified modeling choices,
which depend on the kind of analysis we would like to carry out. 

When choosing the priors, in contrast, we can often rely
on standard recommendations that primarily depend on the domain of each variable.
This does not mean that priors (or likelihoods, for that matter)
can be always chosen blindly using a fixed table of recommendations.
In the following sections, as we go through the various steps
of the Bayesian data analysis workflow, we will validate
our choices of priors and likelihood.
If validation fails, we have to go back and revise the model: priors, likelihoods, or both.

\begin{figure}
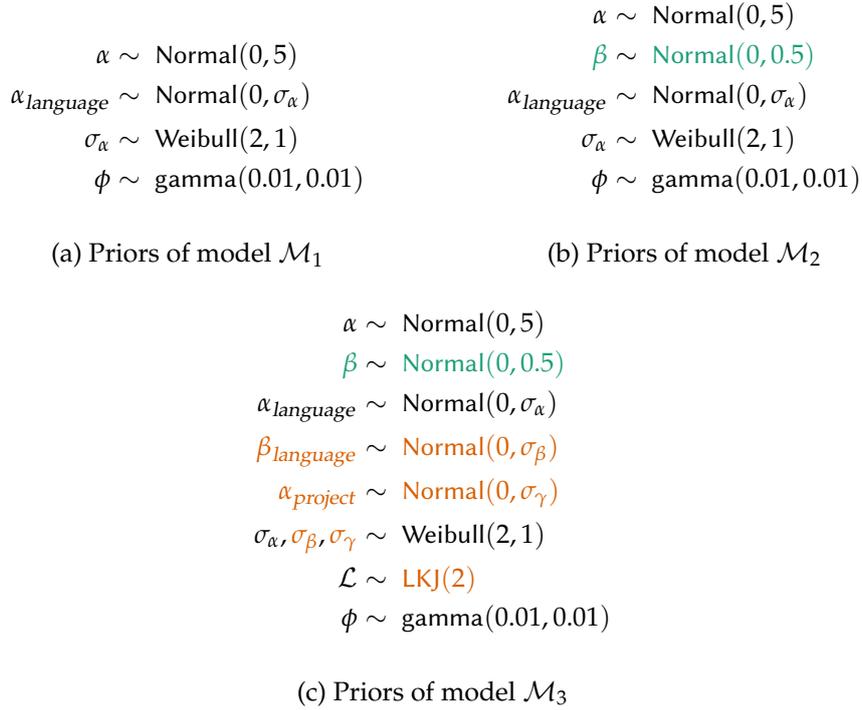

    \centering
    \small
    \begin{subfigure}[b]{0.33\textwidth}
    \centering
    \begin{align*}
    \alpha &\sim\ \dist{Normal}(0, 5)
    \\
    \alpha_{\var{language}} &\sim\ \dist{Normal}(0, \sigma_\alpha)
    \\
    \sigma_\alpha &\sim\ \dist{Weibull}(2, 1)
    \\
    \phi &\sim\ \dist{gamma}(0.01, 0.01)
    \end{align*}
    \caption{Priors of model $\model_1$}
    \label{fig:p1}
    \end{subfigure}
    \hspace{5mm}
    \begin{subfigure}[b]{0.33\textwidth}
    \centering
    \begin{align*}
    \alpha &\sim\ \dist{Normal}(0, 5)
    \\
    {\color{acol} \beta} &\sim\ {\color{acol} \dist{Normal}(0, 0.5)}
    \\
    \alpha_{\var{language}} &\sim\ \dist{Normal}(0, \sigma_\alpha)
    \\
    \sigma_\alpha &\sim\ \dist{Weibull}(2, 1)
    \\
    \phi &\sim\ \dist{gamma}(0.01, 0.01)
    \end{align*}
    \caption{Priors of model $\model_2$}
    \label{fig:p2}
    \end{subfigure}
    \begin{subfigure}[b]{0.33\textwidth}
    \centering
    \begin{align*}
    \alpha &\sim\ \dist{Normal}(0, 5)
    \\
    {\color{acol} \beta} &\sim\ {\color{acol} \dist{Normal}(0, 0.5)}
    \\
    \alpha_{\var{language}} &\sim\ \dist{Normal}(0, \sigma_\alpha)
    \\
    {\color{bcol} \beta_{\var{language}}} &\sim\ {\color{bcol} \dist{Normal}(0, \sigma_\beta)}
    \\
    {\color{bcol} \alpha_{\var{project}}} &\sim\ {\color{bcol} \dist{Normal}(0, \sigma_\gamma)}
    \\
    \sigma_\alpha, {\color{bcol} \sigma_\beta},
    {\color{bcol} \sigma_{\gamma}}
    &\sim\ \dist{Weibull}(2, 1)
    \\
    \lcal &\sim\ {\color{bcol} \dist{LKJ}(2)}
    \\
    \phi &\sim\ \dist{gamma}(0.01, 0.01)
    \end{align*}
    \caption{Priors of model $\model_3$}
    \label{fig:p3}
  \end{subfigure}
  \Description{Equations describing the priors of $\model_1$, $\model_2$, and $\model_3$ given as text.}
    \caption{The priors of statistical models $\model_1$, $\model_2$, and $\model_3$.
    Colors highlight the terms that are added to each model compared to the previous ones.}
    \label{fig:priors123}
\end{figure}

\paragraph{Model $\model_1$.}
As shown in \autoref{fig:p1}, 
we use weakly informative priors for model $\model_1$ 
that are based on the normal distribution.
As we will see during the plausibility analysis (\autoref{sec:plausible}),
model $\model_1$ is so simplistic that its performance is not affected much by the
choice of priors; nonetheless, we discuss its priors in some detail
because we will build on them to choose priors for the more complex models.

The intercept $\alpha$'s prior has mean $0$ 
(that is, we do not know a priori whether the intercept
is positive or negative) and standard deviation $5$.
Remember that the estimated parameter $\lambda$ is log-transformed (see \autoref{fig:m2});
therefore, we can appreciate how weakly constraining this prior is:
two standard deviations on each side of zero span the interval from $e^{-10} \simeq 0$ to $e^{10} \simeq \numprint{22000}$
on the bug counting scale;
that is, the prior only assumes that a project's bugs are up to \numprint{22000} with
95\% probability---which is not a strong assumption at all~\cite{scholzT20LOCs}.
Besides, a normal distribution has \emph{infinite} support, and hence
it does not rule out any count of bugs if the data provides evidence for it.
The prior for the language-level intercept $\alpha_{\var{language}}$ is also a normal distribution with mean $0$; however, choosing the same standard deviation
$\sigma_\alpha$ for every language would 
defeat the purpose of having language-level intercepts.
Instead, we let $\sigma_\alpha$ be a random variable, and assign
a prior to it.
Distributions with support limited to positive values are suitable
priors for standard deviations---which must be nonnegative values.
In this case, we use a \dist{Weibull} for $\sigma_\alpha$
and the default \dist{Gamma} for the dispersion parameter $\phi$ of the negative binomial.

\paragraph{Model $\model_2$.}
In addition to $\model_1$'s weakly informative priors
for $\alpha$, $\alpha_{\var{language}}$, and $\phi$,
$\model_2$ needs a prior for the slope parameter vector $\beta$.
Here too we use a simple normal distribution with mean $0$ (so that 
there is no bias in the possible direction of each predictor's effect)
and standard deviation $0.5$ (which still allows for a broad variability on
the logarithmic scale).
The $\beta$'s prior standard deviations are smaller than the $\alpha$'s
because $\alpha$ %
determines the population average, 
and then $\beta$ moves this average according to each predictor's effect (which needs only introduce a smaller variation relative to the average). \autoref{fig:p2} shows the overall priors for $\model_2$.

\paragraph{Model $\model_3$.}
We choose the prior for the new part of $\model_3$---the language-level slopes $\beta_{\var{language}}$---similarly to how we chose the language-level intercepts $\alpha_{\var{language}}$: a normal with mean 0 and 
a random variable $\sigma_\beta$ for standard deviation.
However, there is an additional technicality that we need to handle:
vectors $\alpha_{\var{language}}$ and $\beta_{\var{language}}$
are not independent but are components of a single \emph{multivariate}
normal distribution with a variance \emph{matrix}~$S$.
Variance matrix~$S$ combines diagonal matrices with the components of $\alpha_{\var{language}}$ and $\beta_{\var{language}}$  
and a covariance matrix $\lcal$.
The customary prior for covariance matrices is a multivariate Lewandowski-Kurowicka-Joe 
distribution; 
$\dist{LKJ}(2)$ is a weakly
informative prior using this distribution, which assigns low probabilities
to extreme correlations.
Finally, the project-level intercept $\alpha_{\var{project}}$'s
prior is also a normal with mean 0 and a random variable $\sigma_\gamma$ for standard deviation.
Just like for the other standard deviations, we choose a a \dist{Weibull}
as prior distribution of $\sigma_\gamma$---a weakly informative distribution
that constrains $\alpha_{\var{project}}$'s standard deviation to
be a non-negative value. %
\autoref{fig:p3} shows the overall priors for $\model_3$.

\begin{figure}
    \centering
    \begin{subfigure}{0.8\textwidth}
      \includegraphics[width=\textwidth]{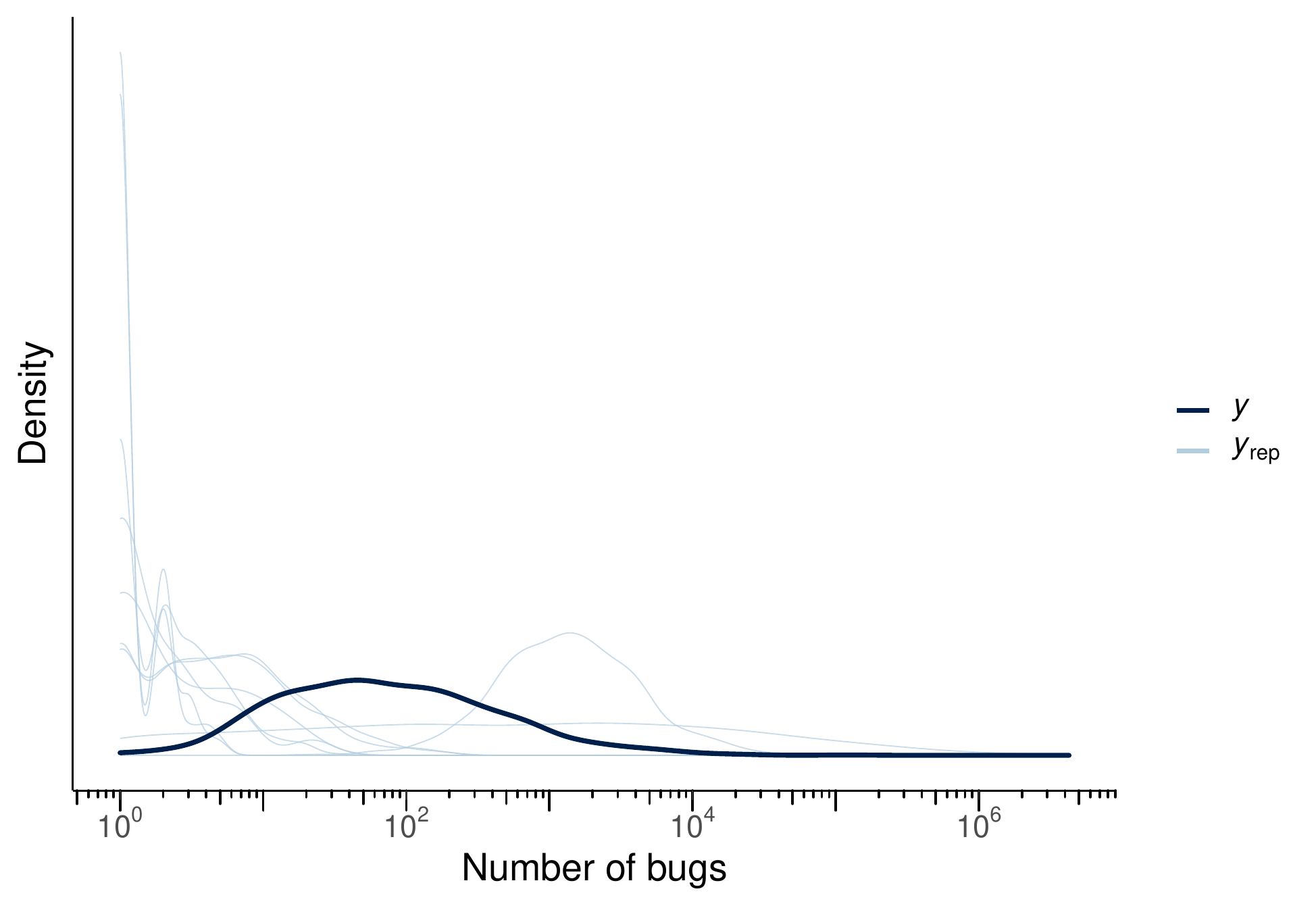}
      \Description{Plot showing several simulated distributions of $\model_2$'s prior, with wide supports spanning all values of the $x$ axis (number of bugs) up to one million ($10^{6}$).}
      \caption{Prior predictive simulation for $\model_2$.}
      \label{fig:ppc-m2}
    \end{subfigure}
    \begin{subfigure}{0.8\textwidth}
      \includegraphics[width=\textwidth]{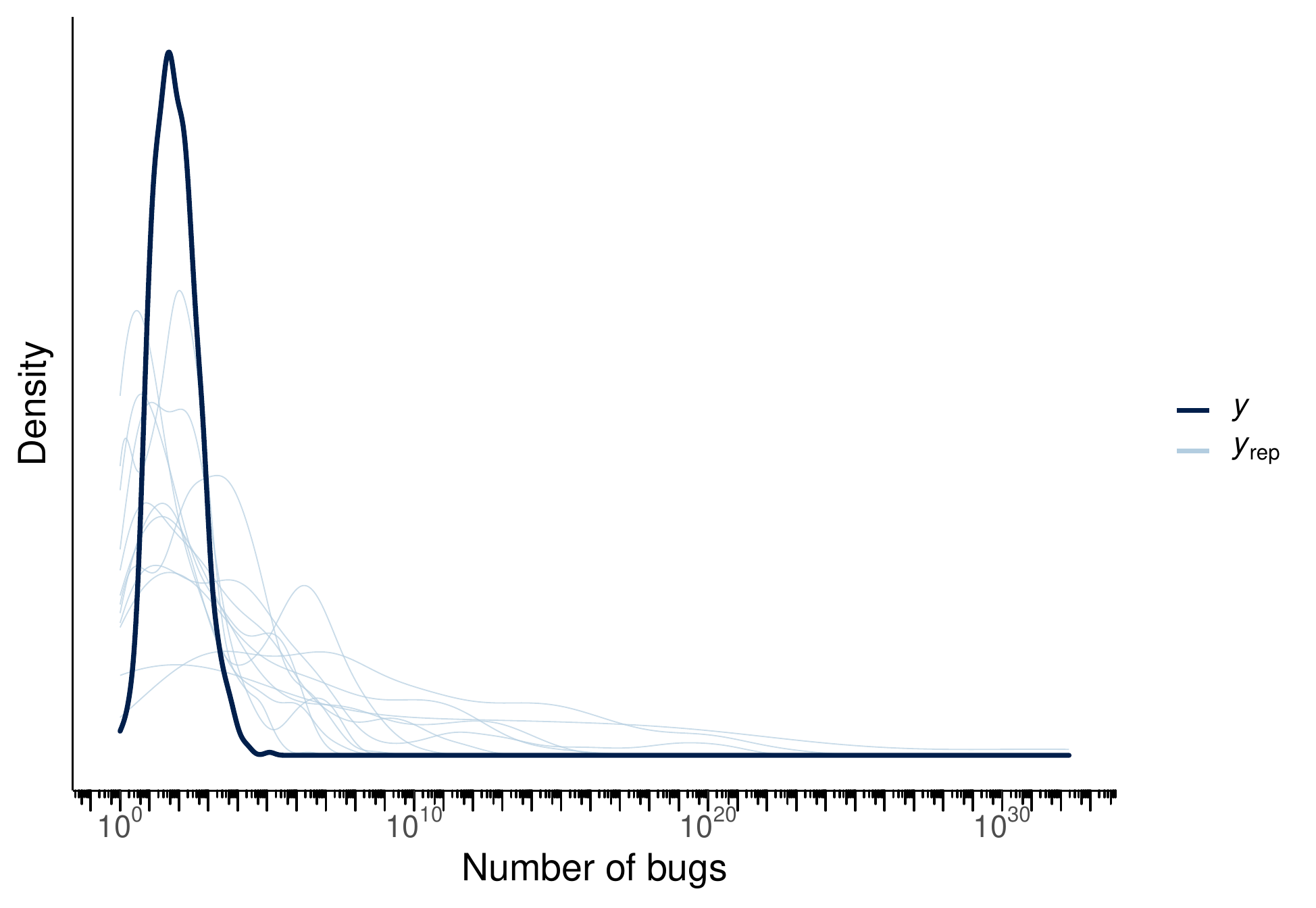}
      \Description{Plot showing several simulated distributions of $\model_3$'s prior, with wide supports spanning all values of the $x$ axis (number of bugs) up to one nonillion ($10^{30}$).}
      \caption{Prior predictive simulation for $\model_3$.}
      \label{fig:ppc-m3}
    \end{subfigure}

    \caption{Prior predictive simulation plots for models $\model_2$ and $\model_3$: each thin light blue line pictures one simulated distribution of the number of bugs in a project drawn from the priors. For comparison,
    the thick dark blue line pictures the distribution of the number of bugs in the measured data. The horizontal scale is logarithmic in base $10$.}
    \label{fig:ppc}
\end{figure}

\subsection{Plausibility}
\label{sec:plausible}

The prior predictive checks are straightforward for all three models,
confirming that our choice of priors---based on standard recommendations for
these kinds of models---leads to plausible outcomes.
As an example, \autoref{fig:ppc-m2} shows several distributions of the outcome variable 
\var{bugs} obtained
with prior predictive simulations of $\model_2$.
These span a very wide support that goes from zero\footnote{The logarithmic link function guarantees a lower bound of zero.} up to over a million bugs per project.
While there are no theoretical limits on the number of bugs in a project independent of its size, %
it is realistic that most projects have less than one million \emph{known} bugs,
and the majority of projects have less than a few thousands---simply because
not many projects have more than one known bug for each line of code~\cite{scholzT20LOCs},
and hence 
a project's size in lines of code is a workable upper bound on the number of distinct bugs.
Anyway, the priors still allow even larger bug counts, but assign
to them increasingly smaller probabilities.
\autoref{fig:ppc-m2} also displays the empirical distribution of bug counts
in FSE's dataset (thick dark blue line); this visually confirms
that the priors are not too restrictive and reflect reasonable expectations.
The prior predictive checks of model $\model_3$ is shown in \autoref{fig:ppc-m3},
and leads to qualitatively similar conclusions---in fact even stronger, given that the priors stretch past an astronomical number of bugs---about the model's plausibility.
So do the prior predictive checks of model $\model_1$, which we do not show %
for brevity.

\subsection{Workability}
\label{sec:workable}

\autoref{sec:workable-def} outlined \emph{simulation-based calibration} and Hamiltonian Monte Carlo 
validation metrics to assess a model's workability.
We use the latter, which are extensively supported
by \textsf{Stan}, to determine whether the
sampling process for each of the three models 
reached a stable state.

$\widehat{R}$---the ratio of within-to-between chain variance---is $< 1.01$ for all three models;
this indicates that the chains have converged towards a stationary posterior probability distribution.
The \emph{effective sample size} is at least $0.11$, $0.17$, $0.13$ for all parameters in each of the three models, and confirms that sampling effectively converged.
Fitting all three models does not run into any \emph{divergent transitions}, and 
the trace plots of the models (included in the replication package) look well-mixed.
In summary, all three models work well computationally.

\begin{figure}
    \centering
    \includegraphics[width=0.9\textwidth]{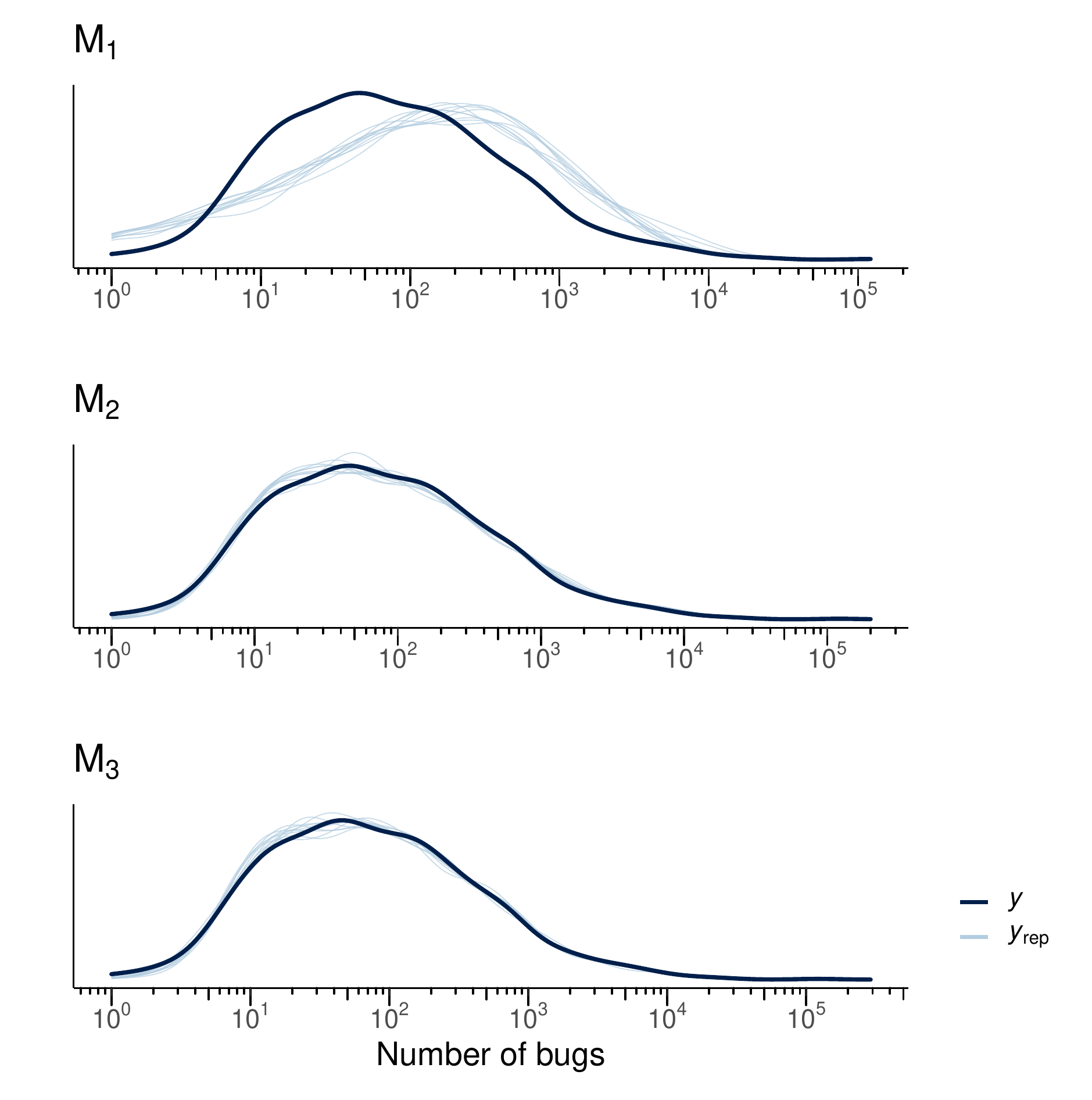}
    \Description{Three plots showing several simulated distributions of $\model_1$'s, $\model_2$'s, and $\model_3$'s posteriors.
      $\model_1$'s posterior tends to outline a bell-like shape with maximum at around $x = 200$ (number of bugs). On the same posterior plot, the actually observed distribution has a maximum at around $x = 70$, which is clearly shifted to the left of the simulated ones.
      $\model_2$'s posterior tends to outline a bell-like shape with maximum at around $x = 70$ (number of bugs). On the same plot, the actually observed distribution follows the same outline as the simulated ones.
    $\model_3$'s posterior looks similar to $\model_2$'s.}
      \caption{Posterior predictive checks for $\model_2$.}
    \caption{Posterior predictive checks plots for models $\model_1$, $\model_2$, and $\model_3$:
      each thin light blue line pictures
      one simulated distribution of the number of bugs in a project
      drawn from the posterior. For comparison,
      the thick dark blue line pictures the distribution of the number of bugs
      in the measured data.
      Model $\model_1$ fails the check because the simulated distributions deviate
      substantially from the data's; in contrast, $\model_2$ and $\model_3$ pass the check
      because the simulated distributions are similar to the data's.
      The horizontal scale is logarithmic in base $10$.}
    \label{fig:postpc}
\end{figure}

\subsection{Adequacy}
\label{sec:adequate}

Let us first study the \emph{adequacy} of our models
with posterior predictive simulations:
we visually compare the distribution of number of bugs per project
in our dataset to several simulated distributions using the fitted models.
The top plot in \autoref{fig:postpc} indicates that $\model_1$ is \emph{not} adequate:
it is too simplistic to capture the data's features; in particular, the
means of the simulated distributions are more than ten times larger than the mean of the data (thick dark blue line).
In contrast, $\model_2$ passes this adequacy test: the middle plot in \autoref{fig:postpc}
shows that the model's predictions look similar to the data.
Model $\model_3$ also passes the visual adequacy test based on the 
posterior predictive simulations, as shown by the bottom plot in \autoref{fig:postpc}.
In \autoref{sec:modeling-2}, we will discuss what these adequacy results tell us about
the interplay between model features and data.

\subsection{Model comparison}
\label{sec:compare}
It is now clear that $\model_1$ is too simplistic, but how to choose between
$\model_2$ and $\model_3$?
Information criteria, which measure the relative adequacy of different models 
fitted on the same data, can help answer this question.
We use the increasingly popular PSIS-LOO information criterion~\cite{vehtariGG17loo},
which works well with models fitted using dynamic Hamiltonian Monte Carlo.\footnote{Compared to more traditional information criteria---such as AIC, BIC, and WAIC---PSIS-LOO can handle non-Gaussian likelihoods (as can WAIC) and also provides diagnostics useful for further analyzing whether a model's posterior behaves well numerically.}
In a nutshell, the criterion ranks the three models
according to their relative adequacy.
It also gives a \emph{difference} score that measures
how well each model performs out-of-sample
predictions relative to the next one in the ranking,
and a \emph{standard error} of the difference, which
quantifies how much more adequate a model is compared to another.
\autoref{tab:loo-differences} displays
the scores for the three models.
Model $\model_3$ is ranked first;
$\model_2$ comes second but its score is 
a whopping 7 standard errors worse than $\model_3$'s;
and $\model_1$ is, unsurprisingly, a distant last.

We conclude that $\model_3$ is the ``best'' model among
the three according to a variety of criteria.
Our analysis will thus use $\model_3$---starting with
a discussion of its features from the point of view of the analysis's goals in~\autoref{sec:modeling-2}.

At some point, one has to stop adding model features and
finalize a model for the current analysis.
Nevertheless, as we discussed in \autoref{sec:uniqueness-optimality},
statistical modeling is never really done:
as more insights, more data, or new techniques become
available, we could go back to the drawing board
and refine the latest model to better capture all available
information.

\begin{table}
    \centering
    \begin{tabular}{crrr}
    \toprule
      \multicolumn{1}{c}{\textsc{model}}
      & \multicolumn{1}{c}{\textsc{rank}}
      & \multicolumn{1}{c}{\textsc{difference}}
      & \multicolumn{1}{c}{\textsc{standard error}} \\
    \midrule
    $\model_3$   &  1 & --   &    -- \\
    $\model_2$  & 2 & $-172.0$  &    $23.6$\\
    $\model_1$ & 3 & $-1963.0$ &     $57.0$ \\
    \bottomrule
    \end{tabular}
    \caption{Each \textsc{model} is \textsc{rank}ed (from better to worse) according to the PSIS-LOO information criterion. The score \textsc{difference}
    between each model and the immediately better one in the ranking, as well as the \textsc{standard error}
    of such difference, quantify the difference in adequacy between models.
    }
    \label{tab:loo-differences}
\end{table}

\section{Bayesian statistical analysis: Results}
\label{sec:bda-results}

When applied following a structured process---like
the one we described in \autoref{sec:guidelines}---Bayesian statistics 
does not simply produce dichotomous answers to research questions.
The outcome of a Bayesian data analysis is a posterior probability distribution,
which we can probe from different angles to get nuanced answers that apply to specific scenarios.
In this section we are going to do this for our case study.

\autoref{sec:modeling-2}
discusses how the Bayesian data analysis process
guided our choice of \emph{models}:
modeling is a looping process,
whose feedback also informs us about key characteristics of the data
we are analyzing.
Not all data features have the same influence on 
the statistics.
\autoref{sec:data-problems}
illustrates how Bayesian analysis can point
to variables with brittle or negligible predictive power
that may indicate problems in how certain measures were operationalized.
After understanding the features and limitations of the fitted
model,
\autoref{sec:practical-significance}
addresses the original study's research questions
in practical settings;
and \autoref{sec:follow-up} outlines
follow-up studies that could address some of the outstanding limitations.

\subsection{Modeling}
\label{sec:modeling-2}

When following the Bayesian data analysis process
in \autoref{fig:workflow},
statistical modeling is based on principles and on
checks that the models are suitable.
This is in contrast to most frequentist statistical practices,
which are primarily based on rules of thumb, conventions (``recipes''), and generic results, 
but may lack operational model-checking processes
that can assess how much confidence we can put in a certain modeling choice.

\autoref{sec:bda-pl} described such a principled Bayesian modeling process
applied to the programming language data.
The first outcome was ruling out $\model_1$ as inadequate,
which was unsurprising given that $\model_1$ 
ignores most of the information that could explain the dataset's variability.
Still, even checks with predictable outcomes are useful:
if they fail, they confirm that a more realistic model is needed and provide 
a minimal effectiveness yardstick;
if they succeed, they avoid an overly complicated model.
This can be especially important in the context of the software engineering industry,
where a complex model can be more costly to understand, collect data for, and maintain.

The second outcome of \autoref{sec:bda-pl}'s analysis
was indicating that, while both $\model_2$ and $\model_3$
are adequate, $\model_3$ clearly outperforms $\model_2$ in out-of-sample
predictive capabilities.
Informally, this means that $\model_3$ fits the data well,
while still avoiding overfitting.
In other words, $\model_3$'s additional complexity over $\model_2$
is justified by its much better effectiveness.
This outcome is specific to the data that we are analyzing,
and is not something that can be determined a priori for all models.
Iteratively creating multiple models and then comparing them is thus an important part of any analysis.

If we compare the definitions of $\model_2$ and $\model_3$---in particular,
their model specifications in \autoref{fig:m2} and \autoref{fig:m3}---we
can attribute $\model_3$'s superior performance to
its unique features.
Unlike $\model_2$, which only includes population-level effects
that control for project characteristics other than the programming language,
$\model_3$ includes a project-specific intercept
and controls for the same characteristics with language-specific slopes.
Thus, we see that clustering per project \textit{and} per language
captures the dataset's characteristics much better:
if we do not do that, we may lose some of the ``signal'' in the data, or conflate different effects and associate them with a single generic predictor.

An indirect advantage of Bayesian data analysis comes
from the techniques that are commonly used to fit Bayesian models:
flexible algorithmic techniques such as dynamic Hamiltonian Monte Carlo 
that can fit, in principle, models of arbitrary complexity---in contrast
to \textit{ad hoc} frequentist techniques that only work for specific, and often limited,
distributional families.
On the other hand, Bayesian models are often more effective
not simply because they can be more complex.
More complex models invariably fit better, but
unwarranted complexity leads to \emph{overfitting}:
a model fits the data perfectly but fails to generalize.
Bayesian analysis techniques include several features that specifically
limit the risk of overfitting when exploring more expressive models:

\begin{itemize}
\item Multi-level models, such as $\model_3$, introduce \emph{partial pooling},
which smooth\-ens differences between groups of different size.
In our case study, the data about some programming languages is 
more scarce than the data about others.
For example,
only 25 projects use Perl, whereas more than 200 use JavaScript;
thus, overfitting Perl's data is a more serious risk than overfitting JavaScript's.
Partial pooling works by transfering %
some of the information learned by fitting the
larger groups to tune the fitting of the smaller groups,
thus reducing the risk of overfitting the latter (and the whole dataset as a result).

\item Prior predictive simulations---discussed in \autoref{sec:adequate}---check
that the priors we have chosen are \emph{regularizing}:
they are not so constraining that they prevent learning from the data,
but they are also not so weak that they 
cannot prevent overfitting the data.
Being able to choose priors, to select different priors, and to quantitatively
compare their effectiveness is a distinct advantage of Bayesian statistics.
Frequentist statistics usually have flat priors, which are the most
prone to overfitting.

\item The information criteria that we used to select $\model_3$ measure
the out-of-sample prediction performance of one model relative to the others.
Models that are unnecessarily complex will overfit the data, and
hence perform worse predictions for \emph{new} data 
(different from the sample that has been used for fitting).

\end{itemize}

\subsection{Spotting data problems}
\label{sec:data-problems}

The rich information provided by a Bayesian data analysis
may also highlight issues with the quality of (parts of) the data that is analyzed,
and suggest which measures need to be cleaned up or improved.

\paragraph{Measuring size.}
Code size is a basic yet essential measure of complexity,
which correlates with lots of other useful metrics of quality~\cite{GL17-metrics}.
Therefore, controlling for project size is essential when analyzing heterogeneous projects.
To this effect, the FSE study included a variable \var{size} in their
regressive model, which measures the total number of inserted lines in all project commits---and which we called \var{insertions} in our models
to make its actual meaning more transparent.
The TOPLAS analysis criticized this choice of size metric---which does
not take deletions and merges into account---and
reported discrepancies between the raw commit data and the
totals in FSE's dataset.
Does our Bayesian analysis offer any hints about the 
reliability of \var{insertions} as a measure of size?

\begin{figure}
    \centering
    \includegraphics[width=0.7\textwidth]{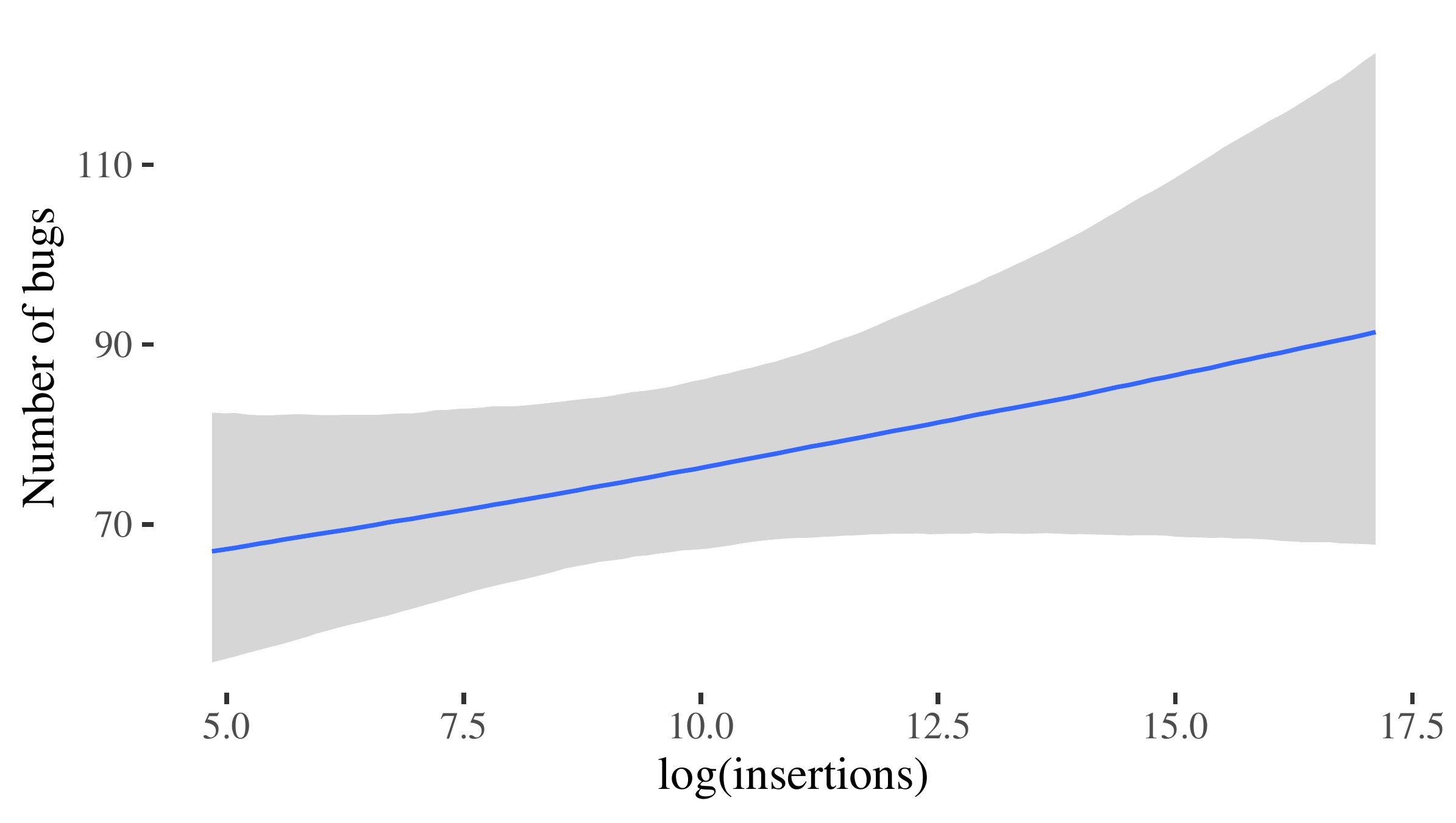}
    \Description{Marginal regression line of number of bugs ($y$) in response to logarithm of number of insertions ($x$). The line goes roughly from $(-5, 70)$ to $(16.5, 90)$. The graph also shows the 95\% error bar on $y$, which is around 27 units to the far left but extends to around 55 units to the far right.}
    \caption{Conditional effects of variable \var{insertions} (the logarithm of the total number of lines added to a project) on the outcome variable \var{bugs} (the number of bugs in a project), corresponding to the marginal distribution derived from the posterior of model $\model_3$.}
    \label{fig:insertions-effects}
\end{figure}

A few results actually single out \var{insertions} as a %
poor predictor compared to the others:
\begin{itemize}
    \item The 95\% probability estimate of its population-level 
    effect includes zero (namely, the (credibility) interval is $[-0.01, 0.06]$), which
    indicates some uncertainty about whether more inserted lines are associated
    with more or fewer bugs on average. 
    Variable \var{insertions}'s mean estimated effect is still positive, but other predictors
    have more clearly defined effects.

    \item The plot of \var{insertions}'s conditional effect on 
    the number of \var{bugs} in \autoref{fig:insertions-effects}
    visually confirms a large uncertainty (again, compared with the other predictors'), 
    which also increases with larger values of insertions.

    \item  The \emph{varying} effect of \var{insertions}
    tend to have larger variance than other predictors---for every language.

    \item If we remove 
    \var{insertions} from $\model_3$, the
    resulting model's predictive performance is practically
    indistinguishable from $\model_3$'s.\footnote{Variable selection~\cite{Bayesian-varselect}---an analysis
      technique that we do not describe in the paper for brevity---also
    suggests to drop variable \var{insertions}. See the paper's replication package 
    for details about this additional analysis of suitability.}
\end{itemize}

All in all, our analysis indicates that \var{insertions}
does not appear to be a particularly useful predictor, and hence
it may not be a reliable measure of code size.

\begin{figure}
    \centering
    \includegraphics[width=0.7\textwidth]{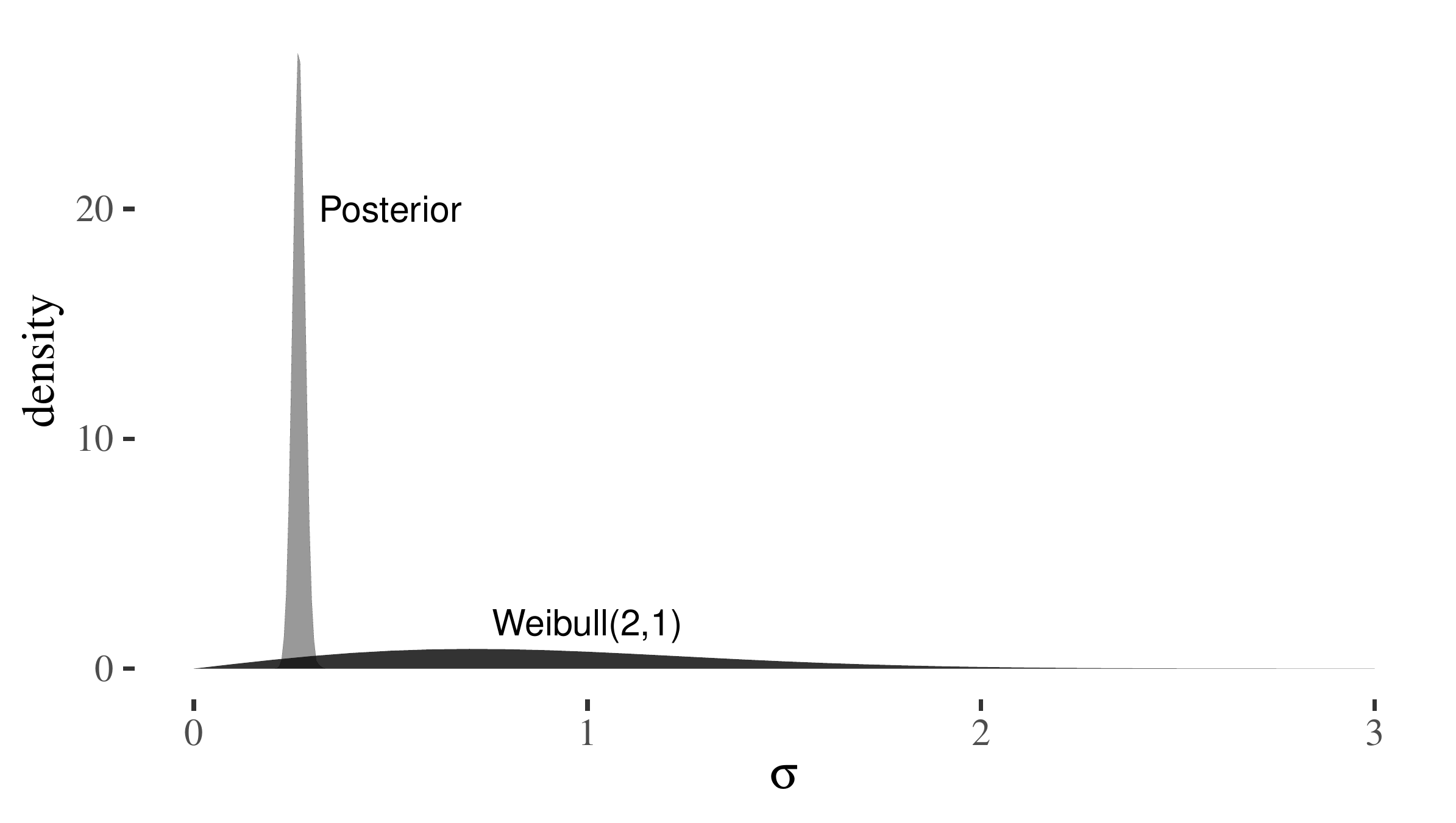}
    \Description{The prior plot is a very wide asymmetric bell-like shape with support from 0 to over 2 ($\sigma$) and maximum value around 0.8 (density).
    The posterior plot is a very narrow spike-like shape with support around 0.25 ($\sigma$) and maximum value close to 30 (density).}
    \caption{Comparison of prior $\dist{Weibull}(2, 1)$ and posterior for project-level intercept $\alpha_\var{project}$'s standard deviation $\sigma_\gamma$ in model $\model_3$. The drastic restriction in uncertainty indicates that the data \emph{swamps} the priors.}
    \label{fig:project-post}
\end{figure}

\begin{figure}
    \centering
    \includegraphics[scale=0.5]{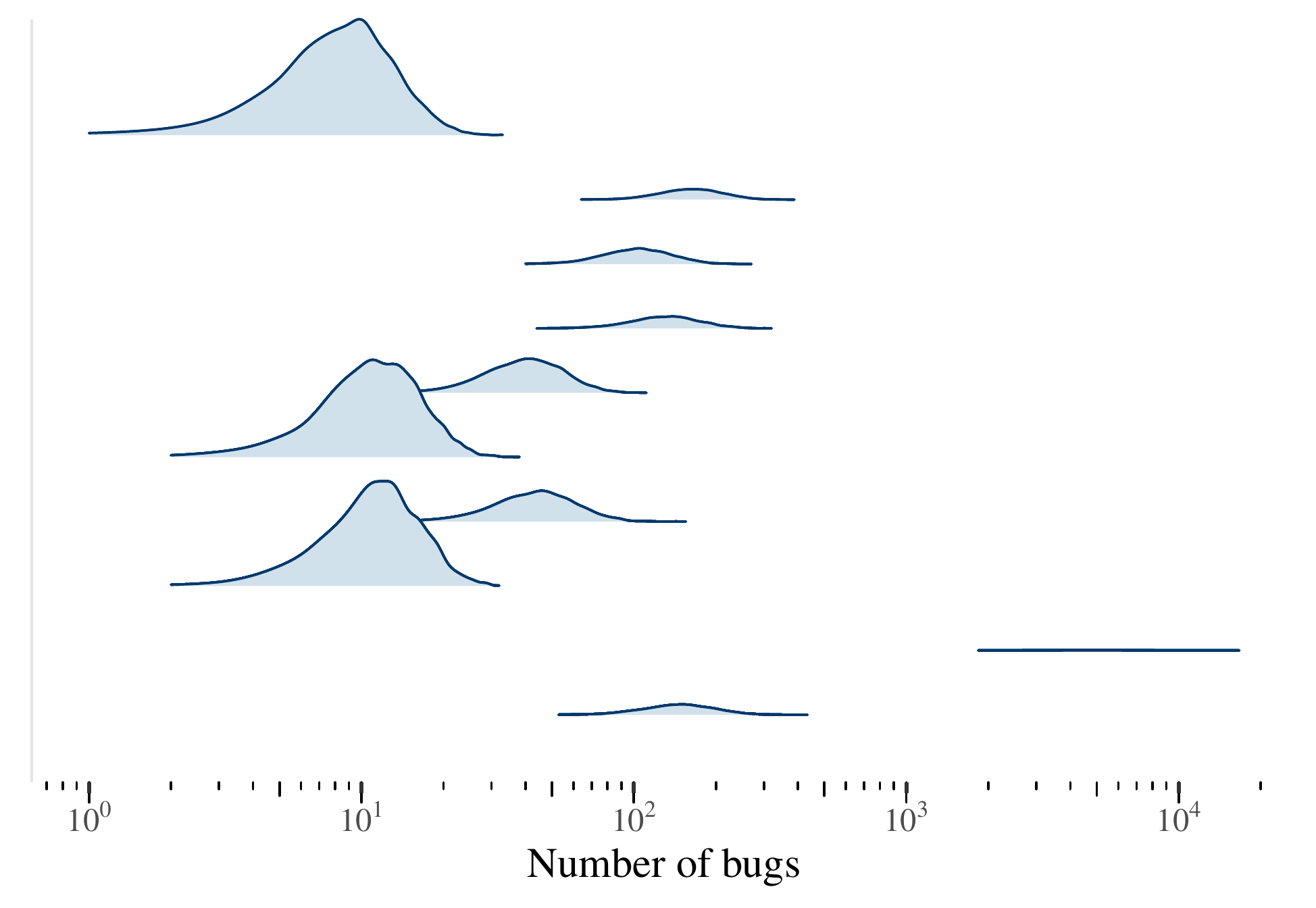}
    \Description{Several distribution of simulated projects, which visibly differ a lot in variance and mean.}
    \caption{Posterior predictions of the bug distributions of $10$ projects
    drawn randomly from the posterior of $\model_3$. The horizontal axis is logarithmic in base $10$.
    The marked differences in shape and location among the distributions
    indicate that projects are heterogeneous.}
    \label{fig:simulate-projects}
\end{figure}

\paragraph{Inter-project variability.}
\autoref{sec:modeling-2} showed that 
a project-specific intercept---which we introduced in $\model_3$---provides better out-of-sample prediction capabilities.
The flip side is that several features of the data vary
considerably from project to project.

The \numprint{1127} data rows are somewhat sparse among the $729$ projects: 
64\% of all projects appear in a single row;
another 26\% in two rows.
Despite these characteristics,
the data \emph{swamps} the priors:
it determines a very precise (that is, narrow) posterior
distribution of the project-specific intercept $\alpha_\var{project}$'s standard deviation $\sigma_\gamma$---shown in \autoref{fig:project-post}.
In other words, the uncertainty about the contribution of 
each project to the overall number of bugs is quite limited:
the data characterizes each project's contribution precisely.
This does not mean that the projects' number of bugs is similar;
on the contrary, the bug distributions of randomly drawn projects
that we get by simulating from $\model_3$'s fitted posterior
differ considerably in shape, support, and mean
(see \autoref{fig:simulate-projects}).
In all, project-specific characteristics 
are an important and well-defined source of information in the data,
which other control variables cannot fully capture.

\subsection{Practical significance}
\label{sec:practical-significance}

Let us now address the original study's research questions.
For brevity, our analysis won't consider 
criteria to classify languages
(``language classes'' such as procedural, functional, scripting, and so on), 
projects
(``application domains'' such as application, database, framework, and so on),
or bugs
(``bug types'' such as algorithm, concurrency, performance, and so on).
These are largely orthogonal to the main focus of the present paper.
Instead, we focus on the key first research question:

\begin{center}
    \textbf{RQ.}~\textit{Are some languages more defect-prone than others?}
\end{center}

Our analysis's information is condensed in the fitted model $\model_3$,
which we can use to generate a distribution
of bugs for every language.
In order to do this,
we have to pick the other inputs of the model:
the number of commits, insertions, age, and developers 
of the hypothetical projects whose number of bugs we are estimating.
While, in principle, these inputs could be any situation or scenario that we want to investigate, it is sensible to start exploring values that are 
close to those observed in the data
used to fit the model 
(following the usual assumption that the sample is representative of the entire population).

\begin{figure}
    \centering
    \includegraphics[width=0.9\textwidth]{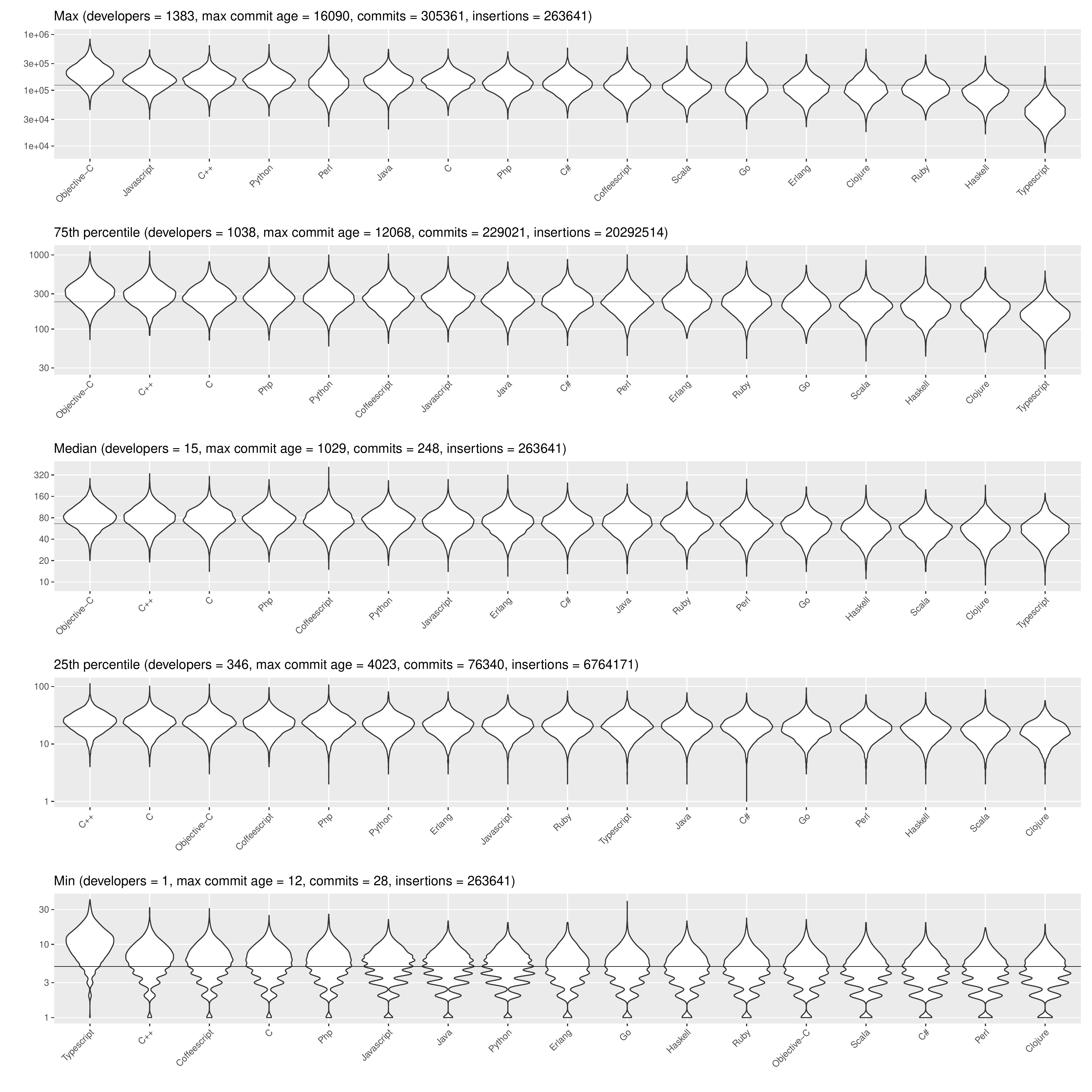}
    \Description{Language order (from most to least error prone) at \emph{maximum} predictors: Objective-C, JavaScript, Python, C++, Perl, Java, C, Php, C\#, CoffeeScript, Scala, Go, Erlang, Clojure, Ruby, Haskell, Typescript. %
      Language order (from most to least error prone) at \emph{75th percentile} predictors: Objective-C, C++, C, Php, CoffeeScript, Python, JavaScript, Java, C\#, Erlang, Perl, Ruby, Go, Scala, Haskell, Clojure, Typescript. %
      Language order (from most to least error prone) at \emph{median} predictors: Objective-C, C++, C, Php, CoffeeScript, Python, JavaScript, Erlang, C\#, Java, Ruby, Perl, Go, Haskell, Scala, Clojure, Typescript. %
      Language order (from most to least error prone) at \emph{25th percentile} predictors: C++, C, Objective-C, CoffeeScript, Php, Python, Erlang, JavaScript, Typescript, Ruby, C\#, Java, Go, Perl, Haskell, Scala, Clojure. %
      Language order (from most to least error prone) at \emph{minimum} predictors: TypeScript, C++, CoffeeScript, C, Php, JavaScript, Python, Java, Erlang, Go, Haskell, Ruby, Objective-C, C\#, Scala, Perl, Clojure.
    }
    \caption{Violin plots of the distributions of number of bugs per project per language, 
      obtained from the posterior of $\model_3$ for five simulated scenarios.
      The plot in each row corresponds to a different scenario: %
      from top to bottom plot, the input variables other than
      \var{language} are set to the empirical dataset's
      \emph{maximum},
      \emph{75h percentile} (3rd quartile),
      \emph{median},
      \emph{25th quartile} (1st quartile),
      and \emph{minimum} values.
      Languages are sorted, left-to-right in each plot,
      by decreasing values of the distributions' medians.
      The vertical axes' scales are logarithmic in base $10$.
      The horizontal line in each plot marks
      the median number of bugs per project across all languages.}
    \label{fig:ranks-max-median-min}
\end{figure}

\subsubsection{Ranking all languages}
\label{sec:practical:rank-all}
\autoref{fig:ranks-max-median-min} displays the distributions as violin plots 
for five combinations of input values:
the dataset's \emph{minimum}, \emph{25th percentile}, \emph{median} (50th percentile), \emph{75th percentile}, and \emph{maximum} number of commits, insertions, age, and developers.\footnote{%
  Unlike \autoref{fig:raw-data}, which plots the raw data, \autoref{fig:ranks-max-median-min}'s simulated projects are directly comparable in terms of defect proneness, as they only differ in the used programming language.
  }
Each plot lists the languages in decreasing order of median predicted number of bugs per project:
from most error prone (left) to least (right).

The plots indicate that the relative ordering of languages
can change conspicuously according to the conditions.
For example, C\#'s defect proneness is average for projects with large or median size and age;
but it becomes better than average for smaller, younger projects.
In contrast, \mbox{C\texttt{++}} is less defect prone only in the largest projects,
whereas it is the most or second most defect prone languages for projects of non-maximal size.
Similarly, the relative rank of some language pairs
varies considerably:
for example, Erlang is less error prone than Go in the largest projects;
the opposite is true in projects of smaller size.

A few languages' ranks fluctuate wildly: Objective-C is among the most defect prone languages
except in small projects, when it is among the least;
TypeScript even goes from least defect prone on large and median projects 
to most defect prone on the smallest projects.
These jumps are so extreme that they may indicate
that the data about these languages is somewhat inconsistent or at least patchwork.
Indeed, TOPLAS's reanalysis reported that only about a third of
the commits classified as TypeScript in FSE's data 
actually included TypeScript code; and 
\citeauthor{FSE}'s extended version~\cite{CACM} 
of their original FSE study
dropped several projects classified as TypeScript.
We did not further look into Objective-C's data, despite its high rank fluctuations, because we wanted to
use the original data without changes.
Nevertheless, this is one clear example
of how Bayesian analysis can help spot data problems, 
and hence bolster better substantiated analyses.
This observation about 
the fickle influence of some languages also
corroborates the evidence that other project-specific characteristics
might weigh comparatively more than 
the used programming language.

\autoref{fig:ranks-max-median-min} also shows that
the bug distributions per language are spread out widely---especially for some languages and
especially for projects that are large and long-running---and their ranges extensively overlap.
The heterogeneity of project-specific characteristics may also contribute to these features;
for example, if projects written in language $X$ tend to be on the large side compared to projects written in language $Y$, the uncertainty in $Y$'s error proneness when used for large projects would dominate the comparison with $X$.
This suggests that the data we analyzed does not warrant 
summarizing the language differences using
a single ranking of defect-proneness.

\begin{figure}
    \centering
    \includegraphics[width=0.8\textwidth]{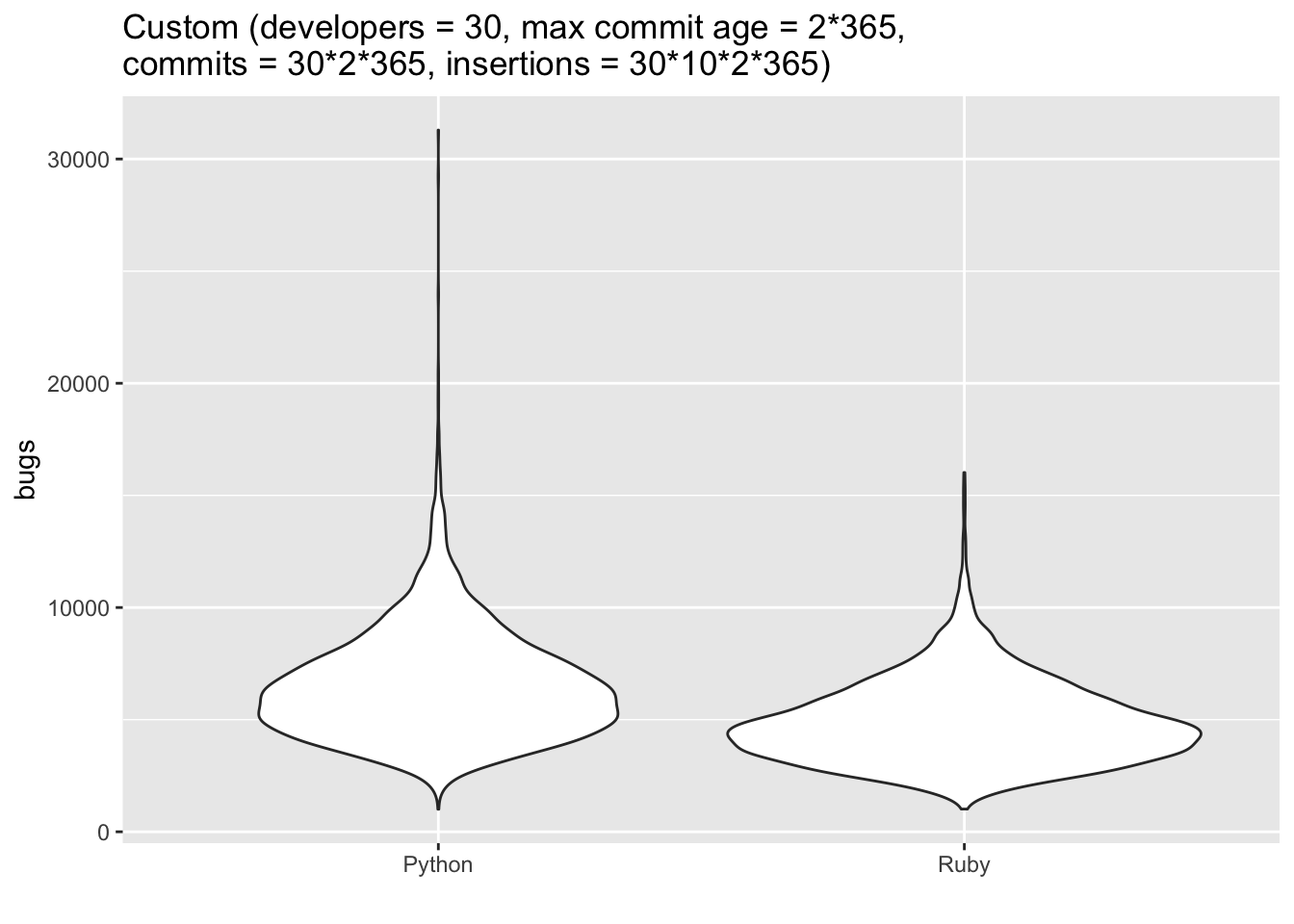}
    \Description{Python's violin plot is widest at around 600 (number of bugs) with a long upward tail that reaches past 30000. Ruby's violing plot is widest at below 500 (number of bugs) with a shorter tail that reaches just 16000.}
    \caption{Violin plots of the bug distributions of Python and Ruby 
    obtained from the posterior of $\model_3$.
    The input variables other than
    \var{language} capture a hypothetical project
    with 30 developers, an age of 2 years, 1 commit per developer every day, 
    and 10 lines added by each developer every day.}
    \label{fig:python-ruby-comparison}
\end{figure}

\subsubsection{Custom scenarios}
\label{sec:practical:scenarios}
While a single ranking of languages according to their absolute defect-proneness
would have little practical meaning,
we can still zoom in on specific conditions that are relevant
\emph{in practice} for a specific project and see what the fitted model can tell us concerning those conditions.

Imagine, for example, we are planning a project that involves around 30 developers
who can code in Python or Ruby;
we estimate the project will run over 2 years, generating an average of 1 commit and 10 lines inserted per programmer per day.
Plugging these numbers 
($\var{developers} = 30$, $\var{age} = 2 \times 365$, $\var{commits} = 30 \times 1 \times 2 \times 365$, and $\var{insertions} = 30 \times 10 \times 2 \times 365$)
into the fitted model $\model_3$,
we get the estimated bug distributions for Python and Ruby 
shown in \autoref{fig:python-ruby-comparison}.
In this scenario, Python tends to be worse (more bugs) than Ruby, since the latter's distribution
has a lower mean, a shorter tail towards high number of bugs, and more mass around lower values.

Whether this evidence is sufficiently strong to decide to choose one language 
over another depends on myriad other factors that are incidental, such as the availability of programmers familiar with one language, the
cost of training new ones, the usability of the programming language
for the project at hand, and so on.
Whatever the practical constraints and requirements may be,
the fitted model can help us meet them by providing
estimates complete with a quantification of their uncertainty.
Realistically, any estimate about the size and development time of a project
is also likely to be somewhat uncertain;
therefore, we would run \emph{multiple} simulations and weigh the evidence summarized by each one against the confidence we have
in the corresponding scenario occurring.

More generally, the results of a principled Bayesian analysis 
facilitate a quantitatively accurate 
transfer of knowledge to practitioners and other researchers. Rather than relying only on overly broad conclusions about the impact of different programming languages, 
using simulations of custom scenarios 
drives follow-up work more precisely:
a practitioner can judge whether the uncertainty in a specific comparison is too large to
base a decision on it; a researcher can decide whether more data is needed to claim more general conclusions.

\subsubsection{Statistical significance}
\label{sec:practical:statisticalsignificance}
Our analysis so far has focused on concrete scenarios defined
in terms of tangible measures in the data domain---such as 
number of bugs and project age.
In contrast, widespread statistical practices (mostly
of a frequentist flavor) try to answer research questions by 
analyzing \emph{statistical significance}, which measures generic 
characteristics of a statistical model.

In a standard regression analysis, 
one usually assesses the \emph{statistical significance} of
each coefficient in the model (also called ``effect'') by checking whether it differs from
zero with a certain probability.
For example, we could compute the distribution of the estimate 
of coefficient $\alpha_{\var{language}}$ for every language.
If $\alpha_{\var{X}}$ is negative with, say, 95\% probability,
we would conclude that language $X$ is associated with fewer bugs
than average with that probability; in other words, $X$ 
is ``statistically significantly'' less error prone than other languages.

FSE and TOPLAS both proceed in such a way,
but using frequentist coefficient estimates 
instead of a posterior probability distribution
on their models---which are similar to our $\model_2$---leading
to their findings about which languages are more error prone than others.\footnote{This is explained in detail in TOPLAS's repetition~\cite[\S~2.2.3]{TOPLAS}, 
which uses FSE's model; TOPLAS's reanalysis~\cite[\S~4.2.1]{TOPLAS} 
suggests a different statistical measure.} 
What about model $\model_3$ fitted on the same data?
The 95\% probability intervals of $\alpha_\var{language}$ include the origin
for \emph{every language}, except TypeScript whose $\alpha_\var{language}$ is strictly positive---but,
as we have commented above, the uncertainty about TypeScript's data puts 
any results about this language on shaky grounds.
Overall, the canonical analysis of statistical significance is just inconclusive
on our model.

To some extent, this outcome is a side effect of $\model_3$'s 
greater complexity over simpler models.
There is a trade-off between the complexity of a model (which
brings greater expressiveness and better predictive performance)
and its interpretability.
The criteria we used to choose $\model_3$ over the simpler $\model_2$
ensure that the former's additional complexity is justified by its much better 
effectiveness.
However, a simple interpretation is no longer feasible:
$\model_3$ includes slope coefficients that also vary with each language,
as well as a project-level contribution;
how each language-specific term interacts with the others is not something
that can be simply estimated with a single coefficient independent of
the predictors' values.

We should appreciate that this is more a feature than it is a limitation.
While mathematically simple models are nice to have,
not all data analysis problems can be addressed with a basic model.
Bayesian analysis techniques do not just support fitting
complex models but provide the means to
handle their complexity and to perform a 
convincing analysis without resorting to
formulaic measures of ``significance''.
The individual model characteristics are not
easy to interpret in isolation,
so that we are forced to interpret the model 
by providing concrete conditions---the number of commits, age, and so on---which 
ground our generic research question onto scenarios
that are realistic and meaningful for our purposes.
In other words, it may be cumbersome to reason about
statistical significance in a Bayesian model but
it is always natural to reason about \emph{practical significance}---which
is what matters most in the end to answer our research questions.\footnote{In related work,
we discuss in greater detail methods to analyze practical significance based on Bayesian data analysis~\cite{TFFGGLE-PracticalSignificance}.}

\paragraph{Model interpretability.}
The focus on practical significance follows from the specific research problem we considered:
answering the question of whether some programming languages are more prone to defects requires precise \emph{predictions}
about the defect-proneness of projects written in different programming languages.
On the other hand, the interpretability of a statistical model may become crucial when targeting other kinds of research questions.
In these scenarios, Bayesian models can still be practically effective.
A relatively complex model like $\model_3$ is not easy to interpret \emph{directly};
that is, we cannot easily and unambiguously assign a direct interpretation to each individual fitted parameter.
However, it is amenable to interpret \emph{indirectly}:
we formulate some scenarios in terms of model variables,
and then we simulate those scenarios on the fitted model.
Interpreting the simulation's results is how we indirectly interpret the model's characteristics.
Ultimately, a key feature of Bayesian data analysis techniques
is what makes these analyses so flexible:
we can combine (indirect) interpretability and predictive capabilities
because the outcome of an analysis is a (sampled) posterior probability distribution---a rich
and actionable source of information.

\subsubsection{Effect sizes}
\label{sec:practical:effectsizes}

\autoref{sec:practical:rank-all} 
demonstrated that the fault proneness of a language over another 
strongly depends on the conditions in which the languages are to be used.
If we have specific scenarios in mind, we can just simulate those
as discussed in \autoref{sec:practical:scenarios}.

Another approach is to compare languages pairwise by %
simulating their performance on a population that resembles the observed data.
Since the comparisons are quantitative---in the form of derived distributions---they can be seen as an effect size, but 
relative to each language pair instead of absolute for all languages at once.

As usual, simulations are derived from the posterior, which
entails that there is no multiple comparisons problem~\cite{miller-MC}:
all information is encoded jointly by the posterior; the pairwise
comparisons are just projections of some of that information.
For the same reason, 
we do not have to commit to a certain way of comparing languages
when we build the model (for example, by choosing how to encode contrasts):
we just select the ``best'' model according to its performance,
and then derive all the information we are interested in from the model
fitted on the data.

\begin{figure}
    \centering
    \begin{subfigure}{\textwidth}
      \includegraphics[width=0.8\textwidth]{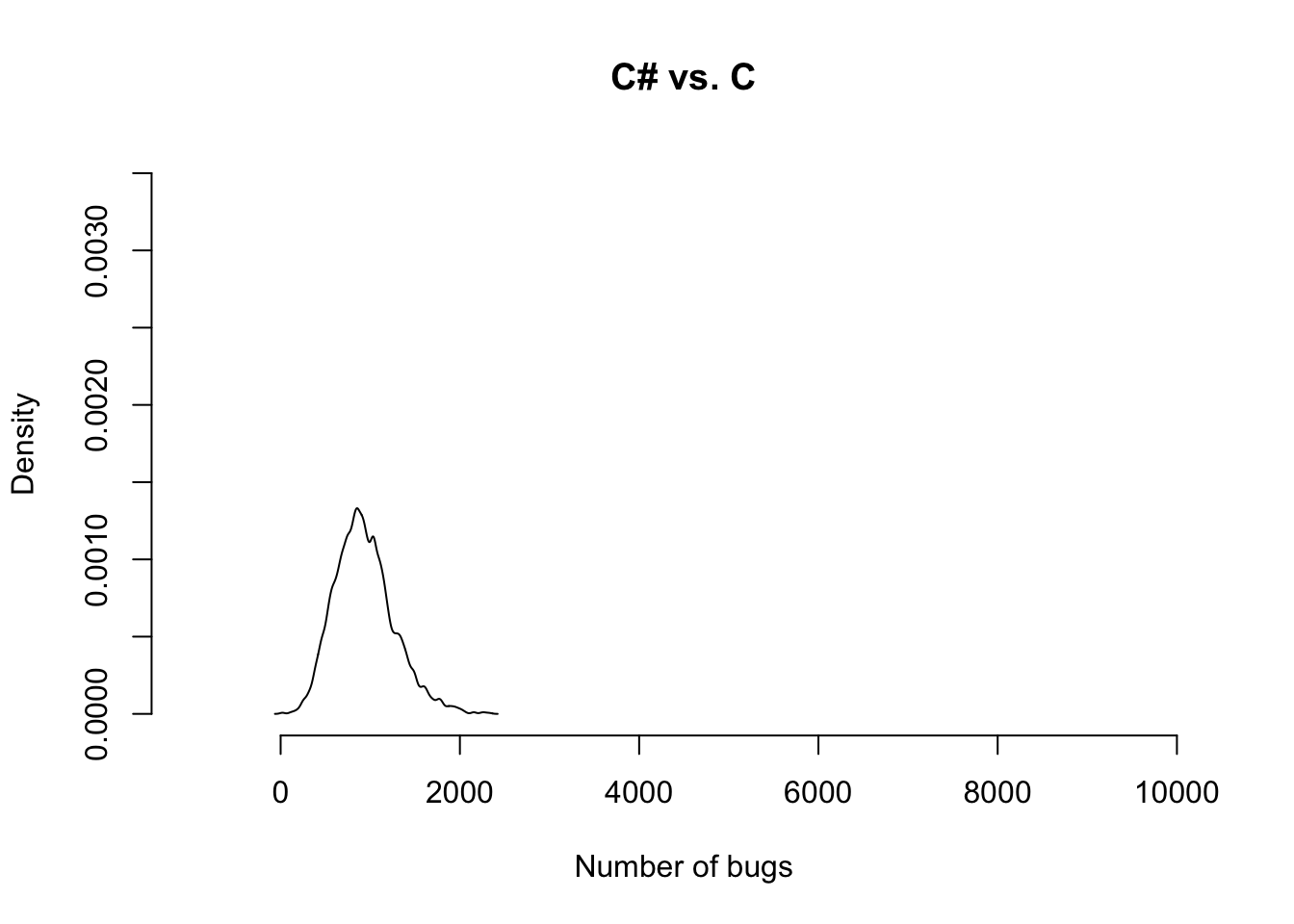}
      \Description{Distribution centered at around 900 (number of bugs) that is entirely to the right of the origin.}
      \caption{Posterior distribution of the difference $\var{bugs}_{\text{C\#}} - \var{bugs}_{\text{C}}$ when predictors are set to the same values as in the empirical data. This indicates that C\# is consistently more fault-prone than C, since it leads to more bugs.}
      \label{fig:es-csharp-c}
    \end{subfigure}
    \begin{subfigure}{\textwidth}
      \includegraphics[width=0.8\textwidth]{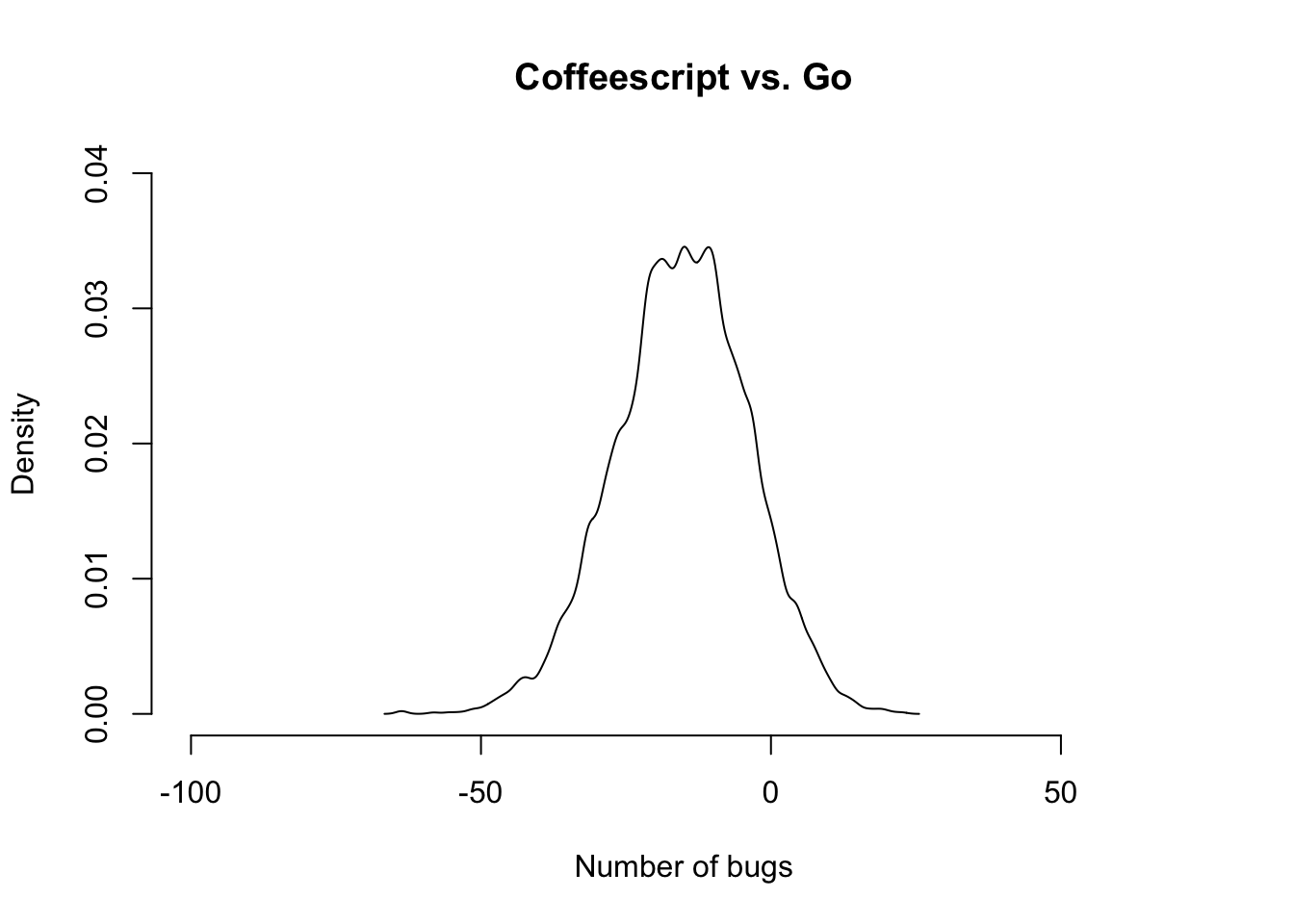}
      \Description{Distribution centered at around $-5$ (number of bugs) that clearly overlaps the origin.}
      \caption{Posterior distribution of the difference $\var{bugs}_{\text{CoffeeScript}} - \var{bugs}_{\text{Go}}$ when predictors are set to the same values as in the empirical data. This indicates that Coffeescript is somewhat less fault-prone than Go.}
      \label{fig:es-coffeescript-go}
    \end{subfigure}
    \caption{Probability distributions of the difference in bug proneness between pairs of languages according to the posterior distribution with population data.}
    \label{fig:effect-sizes}
\end{figure}

Concretely, 
take two languages $\ell_1$ and $\ell_2$ that we
want to compare for bug proneness.
For every data point $d$ in the empirical data,
we set, in the posterior, all predictors except the language to their values in $d$.
Then, we simulate the distribution of %
the expected difference $\var{bugs}_{\ell_1} - \var{bugs}_{\ell_2}$ in bugs 
produced when using one language over the other.\footnote{We can compute the absolute difference in number of bugs because the difference is between samples where the language is the only project characteristic that changes.}

\autoref{fig:effect-sizes} plots the distributions comparing
two pairs of languages, which we selected to demonstrate qualitatively different outcomes of the pairwise comparisons.
The distribution of $\var{bugs}_{\text{C\#}} - \var{bugs}_{\text{C}}$
in \autoref{fig:es-csharp-c} covers only nonnegative values, which
means C\# was consistently more fault prone than C\@.
The distribution of $\var{bugs}_{\text{CoffeeScript}} - \var{bugs}_{\text{Go}}$
in \autoref{fig:es-coffeescript-go}
covers negative values more often than positive ones, denoting that 
Go tended to be more fault prone.
Precisely, we can compute that CoffeeScript was more error prone than Go only around 8.5\% 
of the times.

In the end, we did not really answer the original research question---not 
with a definitive, straightforward answer at least.
Instead, our analysis identified sources
of uncertainty in the data, 
provided means of simulating custom scenarios,
and compared pairs of languages in conditions similar to the collected data's.
This is a solid basis to understand 
what questions can and cannot be answered
by the data, and to plan follow-up data collections and
analyses that zero in on understanding specific outcomes.

\subsection{Planning the next study}
\label{sec:follow-up}

Our analysis %
shed light on the relationship between programming languages
and fault proneness, but also discovered restrictions on how general the findings can be and which factors should be considered.
How can we make further progress in this line of research---beyond
the limitations of what is available in FSE's dataset?

The outcome of our analysis---in particular, the issues discussed in \autoref{sec:data-problems}---help to plan follow-up studies too.
A recurring issue was the clear impact of project-specific characteristics,
which sometimes dominate over language-specific features.
There are at least two ways of better accounting for project features.
One is collecting more data that characterize projects along more dimensions;
for example, a project's domain, the development process it uses, the expertise of its developers, and so on.
The other way is to give up generality and focus on analyzing 
a specific, homogeneous set of projects: the more characteristics are similar
among projects the more accurately the impact of programming languages can be singled out.

When we simulated different scenarios in \autoref{sec:practical-significance}, we found that the \emph{uncertainty}
in the outcome is more pronounced for certain languages than others.
For example, the uncertainty about Objective-C's and Perl's fault proneness is very pronounced on large projects (see the vertical spread in their violin plots of \autoref{fig:ranks-max-median-min}).
To reduce this uncertainty, we should collect more data on large Objective-C and Perl projects, focus on smaller projects, or a combination of both.

More generally, Bayesian models and techniques help zoom out of
each individual study to considering a line of studies in the same subject area.
Each study collects additional data, refines the knowledge that we have of
the area, and identifies further aspects that can be improved---something
follow-up studies will do.
In a way, this realizes a sort of optimization process whose goal is
maximizing knowledge over time.
Bayesian optimization algorithms exist that carry out this process
automatically on a large dataset that can be analyzed incrementally~\cite{ru2018fast};
scientific research deploys processes that do something similar 
on a much longer time scale and with key contributions from human intuition.
The benefits we highlight are not only conceptual; the posteriors from previous studies can be directly used when creating priors for follow-up studies. This can thus enable a more direct and precise way for research studies to build on each other and gradually refine the scientific knowledge.

\section{Related work}
\label{sec:relwork}

We discuss related work in three areas, broadly connected to the paper's contributions:
\autoref{sec:rw-modeling} briefly reviews some widely-used
statistical models other than those we deployed in this paper;
\autoref{sec:rw-guidelines} summarizes other work about statistical analysis guidelines (for frequentist and for Bayesian techniques);
and \autoref{sec:rw-replication} outlines the state of replication studies in
empirical software engineering research.

\subsection{Statistical data modeling and analysis}
\label{sec:rw-modeling}
As we remarked in \autoref{sec:scope}, this paper's guidelines are
largely independent of the specific features of the chosen statistical
models.  All our examples used (generalized/hi\-er\-ar\-chi\-cal) linear
\emph{regressive} models---the workhorse of statistical
analysis~\cite{gelman2020regression}.  Besides their great
flexibility, another practical advantage is that they have
been widely used also with frequentist statistics; therefore, they
offer a convenient bridge for a gradual transition to Bayesian
statistics.

Nevertheless, other classes of statistical models can be used for similar analyses.
One alternative, broad class of statistical models are so-called \emph{graphical} models~\cite{graphical-models},
which use graphs to encode the probabilistic relations between variables.
Bayesian networks~\cite{BN2R} are arguably the best-known kind of graphical models,
which can be used both directly as probabilistic classifiers~\cite{DM-uiqm,codereviews-BN-KrutauzDRM20}
but have also become the basis to encode \emph{causal} relations that go beyond mere correlations~\cite{causality-book}.
The expressive power of Bayesian networks and hierarchical regressive models significantly overlap:
among other things,
one can encode a regressive model as a Bayesian network, and then use network's fitting techniques to analyze it~\cite{R2BN};
conversely, one can encode a Bayesian network as a hierarchical regressive model~\cite[Ch.~5]{BN2R},
and then apply similar analysis techniques as those we demonstrated in the paper.

Machine learning toolkits such as Weka~\cite{weka-book}
provide a convenient way of experimenting with a wide variety of
classical statistical analysis models,
which have been frequently used in the
analysis of software engineering empirical data~\cite{ml1,ml2}
and provide additional serviceable classes of statistical models.
While our paper's guidelines would remain applicable, at least at a high level,
to compare other, widely different statistical models,
doing so in practice may require developing new analysis techniques or
extending existing ones.
In particular, some information criteria---which are used for Bayesian model comparison as discussed in \autoref{sec:comparison-def}---are only applicable to statistical models
that can provide multiple samples from a posterior when the fitted model is used for prediction~\cite{vehtariGG17loo}.
Obviously, generalizing the model comparison criteria (and the other techniques for model analysis) so that they are applicable to all
inductive machine learning approaches falls outside this paper's scope.

It is interesting that several modern machine learning algorithms, such as deep neural networks and active learning,
are applicable both in a frequentist~\cite{active-learning-r2} and in a Bayesian~\cite{activelearning-bayesian,activelearning-bayesian-survey,wang2016towards} context.
Our paper's guidelines could remain broadly useful for frequentist models,
but they do not cover \textit{online} approaches (for example, active learning),
where each iteration of a statistical analysis influences which additional data is collected.
Extending some of our guidelines to online approaches is an interesting direction for future work.
At the same time, the increasingly recognized value of practices such as pre-registered studies~\cite{chambers2015registered}
suggests that the \emph{offline} analysis of fixed, previously collected, datasets
will remain an important and common approach in software engineering empirical research.

\subsection{Guidelines about statistical analysis}
\label{sec:rw-guidelines}

\paragraph{Bayesian data analysis guidelines.}
The last decade's progress in algorithms and tools for Bayesian data analysis has been impressive~\cite{carpenter2017stan,ge2018turing,plummer2003jags,lunn2009bugs} but, without practical support, it is not sufficient 
to promote widespread usage in the empirical sciences.
Recent work about developing guidelines to apply Bayesian data analysis techniques~\citep{gabry2019visualization,schadBV20workflow,gelman20workflow},
which \autoref{sec:guidelines} summarized 
in a form amenable to software engineering empirical research,
has been trying to close this gap.
While these proposals differ in their intended audience and level of detail,
they all build on the basic view~\cite{gelman2004exploratory}
that an analysis should go through multiple models, refine them in several iterations,
and compare them. Then,
\citet{gabry2019visualization} focus on visualization and how to use it throughout a workflow;
\citet{schadBV20workflow}, instead, introduce quantitative checks and illustrate them for a specific scientific area (the cognitive sciences). 
Our guidelines combine elements from both~\cite{gabry2019visualization,schadBV20workflow}
but illustrate them in a way that is amenable to empirical software engineering research practices.
Very recently, \citet{gelman20workflow} presented an early draft of a book that
will further refine some of these Bayesian analysis workflows and guidelines. 
Also very recently, \citet{vandeSchoot2021} published 
an accessible primer on Bayesian statistics and modeling for scientists.

\paragraph{Guidelines on using statistics in empirical software engineering.}
Over the years, several guidelines for using statistics in empirical software engineering
have been pro\-posed---all of them focusing on frequentist statistics, which 
remain the norm in empirical software engineering~\cite{gomes2019evolution}.
\citet{arcuri2011practical} focus on analyzing 
experiments with randomized algorithms, and highlight %
the importance of checking the assumptions of each statistical significance test. They also advocate for extensively using non-parametric statistical tests and effect size measures. \citet{menzies2019bad} catalog ``bad smells'' in data analytics studies %
and discuss remedies to excise them. 
Among the techniques they recommend are up-front power analysis, 
reporting effect sizes and confidence limits, 
and using robust statistics and sensitivity analysis.
A recent literature review of ours~\cite{gomes2019evolution} found 
evidence of a positive impact of such empirical guidelines on
the maturity of statistical practice in empirical software engineering research:
statistical testing, non-parametric tests, and effect sizes have all been increasingly used 
in the field over the last 5--10 years.

\subsection{Replication in software engineering research}
\label{sec:rw-replication}

Recent years have finally seen replication studies become more popular in  software engineering research.
Nevertheless, \citeauthor{da2014replication}'s systematic literature review 
found that internal replications (done by the same authors as the original study) are still much more common than external replications (done by an independent group of authors)~\cite{da2014replication}.
Unsurprisingly, \citeauthor{bezerra2015replication}'s related literature review
found that internal replications are much more likely to confirm the results of the replicated study than external replications~\cite{bezerra2015replication}, 
and used this result to question the value of replications compared to meta-analyses.
Both literature reviews found hardly any examples of \emph{reanalyses} (replications limited to data analysis);
similarly, a taxonomy for replications in software engineering does not explicitly mention reanalysis~\citep{baldassarre2014replication}.

In fact, we tried searching for ``\textsl{reanalysis + software engineering}''
in publication databases and found very few relevant hits---mostly 
papers revisiting qualitative data such as interview transcripts, and reanalyzing 
them to address new questions or theories. As one example, \citet{bjarnason2016theory} developed 
a new theory by reanalyzing interview transcripts from an earlier study of theirs. 
In contrast,
\citet{tantithamthavorn2016comments}
revised a meta-analysis of machine learning in software defect prediction~\citep{shepperd2014researcher} and found that several predictor variables of the original study where co-linear. 
Based on a reanalysis of a subset of the same data,
they also questioned some of the original results and implications. This criticism was later disputed, 
on statistical grounds, by the original study's authors~\citep{shepperd2017authors}. 
Our previous work about using Bayesian analysis in empirical software engineering also performed reanalyses of previous studies using Bayesian  techniques~\cite{FFT-TSE19-Bayes2,TFFGGLE-PracticalSignificance}.\footnote{%
  Our previous work targets various applications of Bayesian statistics such as analyzing practical significance~\cite{TFFGGLE-PracticalSignificance}
and dealing with missing data~\cite{TFF-MissingData20};
the case studied developed there also follow some of the guidelines that we explicitly and specifically present in the present paper.
Therefore, they provide further examples of applications of the guidelines to analyze software engineering empirical data---especially
in their replication packages, since the papers' presentations have a different focus. %
}
Another noticeable external reanalysis is of course \citet{TOPLAS}'s of \citet{FSE}, which we summarized in \autoref{sec:introduction}.

Overall, reanalyses of software engineering data remain
uncommon---especially compared to
other scientific areas where they are widespread 
forms of publication, including
those using Bayesian statistics (for example, in astronomy~\citep{gregory2011bayesian} and medicine~\citep{bath2007can}).

\section{Conclusions}
\label{sec:conclusions}

Reaping the benefits of Bayesian statistics requires 
more than powerful analysis techniques and tools.
In this paper, we presented practical guidelines to
build, check, and analyze a Bayesian statistical model 
that summarize recently developed suggestions brought forward by 
prominent statisticians and
cast them in a format that is amenable to empirical software engineering research.

We then applied the guidelines to analyze a large dataset of GitHub projects
that was previously used to study the impact of programming languages
on code quality~\cite{FSE}.
This study was later criticized by a reproduction attempt that failed to
confirm some of the originally claimed results~\cite{TOPLAS}.
Our reanalysis using Bayesian statistics identified some
shortcomings of the data that also emerged in the reproduction attempt %
(such as the large uncertainty associated with data for programming languages such as TypeScript)
and pointed to other possible effects that were not fully accounted for by
the frequentist models of the previous studies~\cite{FSE,TOPLAS}
(such as the disproportionate differences that are project-specific 
rather than language-specific).
Moving on to the previous studies' main research question 
(``Are some languages more defect-prone than others?''),
our Bayesian model lent itself to evaluating the effect of programming languages
in different concrete scenarios rather than in terms of generic ``statistical significance''.
We found that the impact of programming languages can vary considerably
with other contextual conditions, and hence the original research question
does not admit a simple, generally valid answer---at least not with the analyzed data.

Throughout our reanalysis,
a key advantage of Bayesian techniques
was that they can be used to \emph{quantify}
any derived measures of interest,
as well as the \emph{uncertainty} that comes with each measure.
Such capabilities
are useful not only to infer results in each study,
but also to present and share them in a robust way 
with other researchers and practitioners.
A Bayesian quantitative framework focused on practical significance
can also help plan the next studies in a research area---thus
steadying the long-term progress of
software engineering empirical research and enhancing its broader impact.

\section{Acknowledgements}
The computations were enabled by resources provided by the Swedish National Infrastructure for Computing (SNIC), partially funded by the Swedish Research Council through grant agreement no.\ 2018--05973.
We thank Jonah Gabry for reading an earlier draft of this paper and providing helpful comments.

\ifarxiv
\bibliographystyle{plain}
\else
\bibliographystyle{ACM-Reference-Format}
\fi

\end{document}